\documentclass[fleqn,usenatbib]{mnras}

\usepackage{newtxtext,newtxmath}

\usepackage[T1]{fontenc}

\DeclareRobustCommand{\VAN}[3]{#2}
\let\VANthebibliography\thebibliography
\def\thebibliography{\DeclareRobustCommand{\VAN}[3]{##3}\VANthebibliography}

\usepackage{longtable}
\usepackage{pdflscape}

\usepackage{graphicx}	
\usepackage{amsmath}	

\title[The ubiquity of turbulence in PNe]{The ubiquity of turbulence in the expanding kinematics of the ionized shells of Galactic planetary nebulae}

\author[F. Ruiz-Escobedo, M. G. Richer and J. A. L\'opez]{
Francisco Ruiz-Escobedo,$^{1}$\thanks{E-mail: fdruiz@astro.unam.mx}
Michael G. Richer$^{1}$ and
Jos\'e Alberto L\'opez$^{1}$
\\
$^{1}$Universidad Nacional Aut\'onoma de M\'exico. Instituto de Astronom\'ia.
A.P. 106, 22800. Ensenada, B.C. , M\'exico\\
}

\date{Accepted XXX. Received YYY; in original form ZZZ}

\pubyear{\the\year{}}

\begin{document}
\label{firstpage}
\pagerange{\pageref{firstpage}--\pageref{lastpage}}
\maketitle

\begin{abstract}
We present an analysis of the residual velocities from a sample of 105 Galactic planetary nebulae (PNe), the largest done to date on this subject. The analysis has been carried out with long-slit, high dispersion echelle spectra. The data were drawn from the San Pedro M\'artir Kinematic Catalogue of Galactic Planetary Nebulae.  The residual velocity is identified with turbulence in the plasma and is derived by decomposing the emission line profiles into their structural contributors. Turbulence seems pervasive throughout all the PNe in the sample. We find the values for the residual velocities in the sample to be either transonic or slightly supersonic in the ionized environment. When residual velocities of [\ion{N}{ii}], [\ion{O}{iii}] and \ion{He}{ii} in the same PNe are compared, there is a tendency for the residual velocities of the higher ionized ion to be larger by about 5--10 km s$^{-1}$, indicating that the turbulent structure is larger in the inner zones of the PN. We find in general no clear correlation between the residual velocities and other nebular parameters such as morphology, global expansion velocities, ionization degree and binary cores. The only exception is the case of PNe with H-poor ([WR]-type) central stars, where we confirm previous results that have consistently shown higher residual velocities for  this group of  PNe as compared to those with H-rich central star atmospheres.  Turbulence seems to be a localised, random, dissipative process occurring in the inner sections of the shell and may affect its early evolution.

\end{abstract}




\section{Introduction}\label{sec_introduction}

Planetary nebulae are post-AGB objects that constitute one of the last evolutionary stages of low- and intermediate -mass stars (nominally, $M_i = 1-8\,M_{\odot}$).  These objects exhibit shells of ionized gas expanding around a central star at typical velocities of $\sim 25$ km s$^{-1}$.  
In these systems the evolution of the central star governs the development of the surrounding nebular shell through the evolution of its ionizing radiation field and stellar wind power.  The interacting stellar winds theory of \citet{kwoketal1978} and \citet{Kahn1985} describe the basic dynamic evolution of the nebular shells as due to the interaction of the slow, dense wind emitted while the progenitor star was on the AGB and the much faster and dilute wind that develops as the central star's effective temperature increases.  
The global kinematic components include an ordered velocity field of the nebular shell that typically contributes a velocity gradient \citep{wilson1950}.
Nevertheless, the morphology of the ionized shells and the related velocity fields can become complex since the early stages of development, producing collimated outflows and poly-polar structures; see \citet{Lopez2022} for a recent review on the kinematic structure and morphology in PNe.
Hydrodynamical numerical models clearly reveal that the internal dynamics of ionized nebular shells or regions (PNe, \ion{H}{ii} regions or SNR) give rise to instabilities that develop localised turbulence in areas close to shocks or ionization fronts. There are three different fluid interactions that give rise to instabilities, namely: In the Kelvin-Helmhotz case where different flow speeds across the interface between two fluids produce a velocity shear.  Vishniac instabilities occur in radiative shocks as pressure driven growth in a thin, cold shell of swept-up material. Lastly, Rayleigh-Taylor instabilities are produced as a high-density fluid is supported by  a low-density fluid against gravity or acceleration. All these instabilities often lead to turbulent regions as a secondary effect \citep[see e.g.,][]{toalaarthur2014, garciaseguraetal2022}. 

Observationally, turbulence contributes an additional
area or section of an emission line profile that cannot be accounted for as the result of thermodynamics, the nebula's dynamical structure, instrumental resolution, and seeing.
\citet{sabbadin2008} dedicated a detailed study to turbulence in PNe, explaining in detail how to disentangle the various contributions that broaden the line profiles.  On a quantum mechanical level, there is the natural line width, which produces a negligible broadening of all lines (at the m s$^{-1}$ level), and fine structure, which is relevant for H and He lines ($< 10$\,km s$^{-1}$).  The thermodynamics of the plasma contributes thermal broadening ($\sim 21$\,km s$^{-1}$ for H at $10^4$\,K, $\propto \sqrt{m}$).  The ordered velocity field of the nebular shell typically contributes a velocity gradient \citep{wilson1950}, but additional coherent velocity structure may also be present \citep[e.g.,][]{bryceetal1992}.  The process of observation contributes to further broaden the lines, via the instrumental resolution and the seeing, which can mix signals from different parts of the structure if the seeing is a significant fraction of the object's size, e.g. \citep[]{guerreroetal1998}.  If the observation integrates over a large fraction of an object's spatial extent, it will similarly contribute to a broader line profile. Hence, turbulence contributes a measurable additional velocity width to the line profile that can only be accounted for after estimating the effects of the other processes known to be present (\S\ref{sec_residual_velocity}).  In what follows, we shall use the terms \textit{turbulence} and \textit{ residual velocity} when referring to the physical concept and observational effects, respectively.  We also note that turbulence in planetary nebulae has not been characterized as has been done for \ion{H}{ii} regions or the Earth's atmosphere \citep[e.g.,][]{stull1988,garciavazquezetal2023}. 

Turbulent motions have received limited attention in observational studies, for example, \citet[]{guerreroetal1998} found turbulence contributions of the order of 15 km s$^{-1}$ in the study of the kinematics of the attached shells of a sample of 16 PNe. 
\citet{acker2002} studied the line profiles of 16 [WR]-type and 8 regular planetary nebulae, finding that in the first group they needed an additional velocity component to successfully account for the velocity profile, identifying it with turbulence.  Likewise, \citet{gesicki2003} studied the internal kinematics of 14 PNe reporting the presence of turbulence for those objects in their sample with a [WR]-type stellar core.   \citet{medina2006} studied the kinematics of a sample of 47 PNe out of which 24 of them were of the [WR]-type, finding that the turbulence contribution was enhanced in this group. 
Many bright planetary nebulae subtend small angles on the sky that require extremely high spatial resolution to extract well localised spectra. All studies to date measure the residual line width, after accounting for the known effects that broaden the lines, described above. Typically the residual velocity amounts to $10-30$\,km s$^{-1}$.

Here, we study the line profiles in a large sample of well-resolved planetary nebulae at high spectral resolution.  We measure  the velocity splitting and residual velocity from the blue- and redshifted components from split line profiles of [\ion{N}{ii}] $\lambda \lambda$6583, 6548, [\ion{O}{iii}] $\lambda$5007 and \ion{He}{ii} $\lambda$6560 emission lines, which are representative species of the ionization structure within a PN. We describe our methodology in \S\ref{sec_methodology}.  We generally follow the precepts from \citet{sabbadin2008}, though we restrict our observations to lines of sight through or near the central stars to minimize projection effects.  We also decompose the line profiles into individual components, limiting our study to the main emission component from each side of the nebular shell, thereby limiting the effects of coherent velocity structures in broadening the line profiles.  We present our results in \S\ref{sec_results}, of which the most relevant is the persistent presence of a residual velocity varying from $5-30$\,km s$^{-1}$.  We discuss these results in \S\ref{sec_discussion}, and present our conclusions in \S\ref{sec_conclusions}.

\section{Methodology}\label{sec_methodology}

\subsection{Sample selection}\label{sec_sample_selection}

For this work, we used data from the The SPM Kinematic Catalogue of Planetary Nebulae\footnote{\url{http://kincatpn.astrosen.unam.mx/}} \citep[][henceforth ``the SPM Catalogue"]{lopez2012}. The SPM Catalogue contains high-resolution spectra of about 700 Galactic PNe.  The great majority of these spectra were acquired with the Manchester Echelle Spectrograph \citep[MES,][]{meaburn2003} attached to the f/7.5 focal station of the 2.1 m Arcadio Poveda Telescope at the Observatorio Astron\'omico Nacional on the Sierra San Pedro M\'artir (OAN-SPM), Baja California, Mexico.  The MES is a long-slit, Echelle spectrograph  without a cross-disperser that uses narrow-band filters to isolate single orders.  We use data obtained with the [\ion{O}{iii}] and H$\alpha$ filters, which include the lines [\ion{O}{iii}] $\lambda$5007 and \ion{He}{i} $\lambda$5016 (order 114) and [\ion{N}{ii}] $\lambda\lambda$6458,6583, \ion{He}{ii} $\lambda$6560, \ion{H}{i} $\lambda$6563, and \ion{C}{ii} $\lambda$6578 (order 87), respectively.  The MES spectral resolution depends upon the slit width.  Most of the spectra were obtained using slit widths of 70 $\mu$m and 150 $\mu$m, resulting in spectral resolutions of 6 km\,s$^{-1}$ (R $\sim$ 50,000) and 11.5 km\,s$^{-1}$ (R $\sim$ 26,000), respectively.  These data are complemented with spectra acquired using the UCLES spectrometer at the Anglo-Australian Telescope \citep[AAT,][a conventional echelle spectrograph]{diegoetal1990}.  For these spectra, the spectral resolution is about 9.6 km\,s$^{-1}$ (R $\sim$ 31,000).  In addition, we also used recently acquired MES spectra (June 2023) of Hf\,2-2, NGC 6778 and NGC 6826 that were reduced as described by \citet{lopez2012}.  

Our selection criteria are:
\begin{enumerate}
    \item The objects are spatially resolved.
    \item The spectrograph slit crossed the line of sight towards the central star or near it.
    \item The line profile shows a clear splitting of the emission from the approaching and receding sides of the nebular shell, as in Figure \ref{fig:line_prof_fitting}.
    \item Objects with simple morphologies, elliptical and spherical, were favoured; however, some objects with more complex morphologies were included. In these cases, the line profiles did not include complex substructures, and so allowed us to measure the residual velocities in the main nebular shells.
\end{enumerate}

The first criterion allows us to minimize the effects of image quality, which was typically $\sim 1.5\arcsec$.  This was almost always a small fraction of the object size, which spanned the range from 4.8\arcsec to 270\arcsec\ \citep[Strasbourgh-ESO Catalogue,][]{acker1992}, with a median value of 23\arcsec. Spatial resolution also helps minimize the velocity broadening due to mixing the projections of velocities from different parts of each object.  Although we do not present this result below, we checked whether the angular size correlated with the residual velocity and found no correlation.  Hence, we expect that our residual velocities are not biassed by projection effects.  The second criterion helps minimize the effects of projection on the velocity structure since the motion on the line of sight towards the central star should be mostly along that line of sight \citep{sabbadin2008}. The third criterion allows us to use one methodology for all objects and avoids possible degeneracies when fitting multiple Gaussians to an undivided line profile. The last criterion allows us to avoid cases where the velocity field is especially complicated and might lead us to over-estimate the residual velocity.

Our final sample consists of 105 objects.  We list them in Table \ref{tab:longtab_caract}, where we indicate the common and PN G names, the filters used, the nebular morphology \citep{lopez2012}\footnote{The options are bubble (41 PNe), elliptical (29 PNe), bipolar (17 PNe), toroid (7 PNe), compact (7 PNe) and cylinder (4 PNe)}., the central star's spectral type \citep{weidmann2020}, whether it is known to be binary \citep{jones2017, bofin2019, weidmann2020}\footnote{\url{https://www.drdjones.net/bcspn/}, last updated on 8 October 2025}, and the presence of nebular \ion{He}{ii} $\lambda$6560 emission in the SPM Catalogue spectrum.  For those objects for which there are observations at different position angles (PA) for the spectrograph slit, we use the observations that yield the smallest line width, since our basic intention is to establish lower limits for the residual velocities.  The 5 objects for which the spectrograph slit crossed close to the line of sight towards the central star are indicated with an asterisk in Table \ref{tab:longtab_caract}).  103 objects have spectra in the H$\alpha$ filter and 35 in the [\ion{O}{iii}] filter, with 33 having spectra in both filters.  

We focus our analysis on 4 emission lines: [\ion{N}{ii}] $\lambda \lambda$6583, 6548, [\ion{O}{iii}] $\lambda$5007 and \ion{He}{ii} $\lambda$6560.  For these lines, we adopt the following rest wavelengths, $\lambda_0$,: [\ion{N}{ii}] 6583.45, 6548.05, [\ion{O}{iii}] 5006.843, (from NIST\footnote{\url{https://www.nist.gov/pml/atomic-spectra-database}, last updated on November 2024.}, \citealt{kramida2024}) and \ion{He}{ii} 6560.10 \AA \, (from \citealt{vanhoof2018}\footnote{\url{https://linelist.pa.uky.edu/newpage/}, last updated on October 2023.}).  The [\ion{N}{ii}] $\lambda \lambda$6583, 6548, [\ion{O}{iii}] $\lambda$5007 and \ion{He}{ii} $\lambda$6560 lines have ionization potentials that span the full ionization structure within the PNe.  Although the H$\alpha$ line is present in all spectra, we excluded it from our analysis.  The entire nebular volume emits H$\alpha$, so it is the line most affected by any coherent kinematic sub-structure that is present, such as velocity gradients.  This is particularly problematic due to its large thermal broadening that washes out the evidence of this sub-structure, a result of its small mass.  The \ion{C}{ii} $\lambda$6578 and \ion{He}{i} $\lambda$5016 lines were observed each in four objects, so we do not consider them further, but we include the data in Table \ref{tab:longtab_caract}.  

Our selection criteria 1, 3, and 4 as well as the lines we consider and our measurement methodology (following section) are all designed to limit our sensitivity to the effects of large-scale coherent velocity structures, such as velocity gradients.  We consider only the main emission component from either side of the nebular shell, so our residual velocities pertain to rather localized emission regions within the volume of the nebular shell.  In addition, at least the [\ion{N}{ii}] and \ion{He}{ii} lines do not arise throughout the entire volume of the nebular shell, but rather at the inner and outer edges, further restricting the volume we consider in these lines.  (The [\ion{O}{iii}] $\lambda$5007 line could well arise from the majority of the nebular volume.)

\subsection{Line profile fitting}

\begin{figure}
    \centering
    \includegraphics[width=0.85\linewidth]{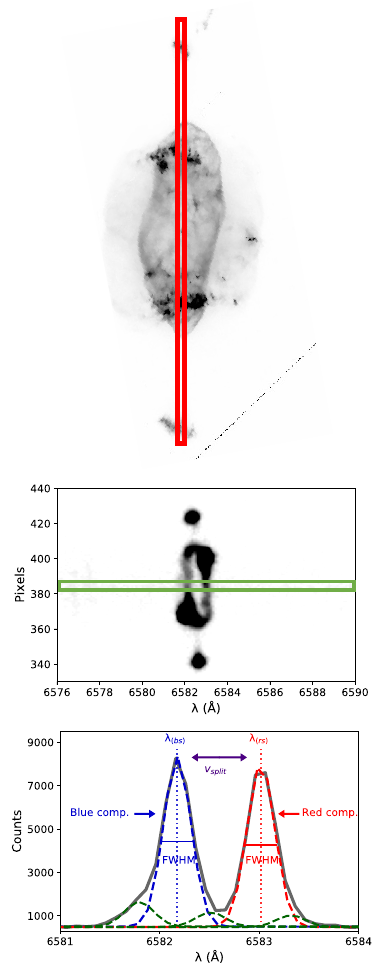}
    \caption{We illustrate our methodology using NGC 7009 as an example \citep[e.g.,][]{richer2025}.  Top panel:  We superpose the MES slit on an HST image ([\ion{N}{ii}] $\lambda$6583; Prop. ID 11122, PI Balick), with the slit crossing over the central star. Middle panel:  We present the 2D spectrum showing the spatial distribution of [\ion{N}{ii}] $\lambda$6583 emission as function of wavelength.  The green band represents the extraction window where we extracted the one-dimensional spectrum shown in the lower panel.  In addition to the blue- and redshifted components, other velocity components are evident.  Lower panel:  We fit Gaussian profiles (dashed lines) to the extracted line profile.  The Gaussian profiles in blue and red indicate the nebular shell's main blue- and redshifted components, respectively, whose FWHM we use to measure the residual velocities.  The Gaussian profiles for the other components that we fit are shown in green.  The dotted vertical lines indicate the central wavelengths of blue- and redshifted components ($\lambda_{(bs)}$, $\lambda_{(rs)}$) that we use to determine the velocity splitting.  
    }
    \label{fig:line_prof_fitting}
\end{figure}

Figure \ref{fig:line_prof_fitting} illustrates the acquisition and format of the data in the SPM Calalogue, as well as the process of fitting the line profiles. The SPM Catalogue contains bidimensional spectra (FITS format).  We extract one dimensional spectra using \textsc{iraf}'s\footnote{\textsc{iraf} is distributed by the National Optical Observatory, which is operated by the Associated Universities for Research in Astronomy, Inc., under contract to the National Science Foundation. 
} \textit{splot}  task  \citep{tody1986, tody1993, fitzpatricketal2024}.  We extracted the spectra at the spatial coordinate of the central star, or at the spatial centre for objects with no detected central star. 
We used the smallest aperture that was sufficiently large to obtain line profiles with adequate signal-to-noise ratio that allowed fitting reliably the emission components that were present in the two-dimensional spectra.  In no case did this aperture exceed 40\% of the object diameter.

As illustrated in the bottom panel of Figure \ref{fig:line_prof_fitting}, we fit Gaussian functions to each line profile in order to separate their different components and measure their FWHM.  The number of components that we fit is based upon both the shape of the one-dimensional line profile and the line structure visible in the two-dimensional spectrum.  Of particular interest here are the main blue- and redshifted components emitted by the approaching and receding sides of the nebular shell, and not its entire emission because our objective is to determine the minimum residual velocity present.
We characterize the kinematics of the nebular shells using the velocity splitting of the two main components and the broadening of the approaching and receding components of the nebular shell.  We define the velocity splitting as half the difference in the radial velocities of the blue- and redshifted components due to the Doppler effect.

\subsection{The velocity splitting and the residual velocity}\label{sec_residual_velocity}

Our main interest is to study the velocity widths of the main blue- and redshifted components emitted by the nebular shells in our sample.  In particular, we wish to understand whether the observed widths are those expected or whether there exist systematic deviations from expectations.  We therefore decompose the measured line profiles to account for the known broadening mechanisms already discussed in \S\ref{sec_introduction}.  To do so, we define a residual velocity, $\Delta W_{ks}$, which is the difference between the observed line width and the sum of the broadening due to the instrumental resolution, thermal motions, and the fine structure of the lines (\ion{He}{ii} $\lambda$6560 only), i.e.,

\begin{equation} 
    \Delta W_{ks}^{2} = FWHM_{obs}^{2} - FWHM_{inst}^{2} - FWHM_{th}^{2} - FWHM_{fs}^{2}
    \label{eq:width_correction}
\end{equation}

\noindent where $FWHM_{obs}$ is the observed width of blue- or red-shifted component measured from the spectra, $FWHM_{inst}$ is the instrumental broadening, $FWHM_{th}$ is the thermal broadening, and $FWHM_{fs}$ is the fine structure width for \ion{He}{ii} $\lambda$6560 \citep[4.2 km s$^{-1}$,][]{sabbadin2008}.  For the instrumental broadening, we measure the width of lines in the arc lamp spectra close in wavelength to the lines of interest (values of 7 -- 14 km s$^{-1}$).  For the thermal broadening, we use \citep{sabbadin2008}:

    \begin{equation}
        FWHM_{th} = 0.215 \sqrt{\frac{T_{e}}{m_{ion}}}\  (\rm{km \, s}^{-1}) 
    \label{eq:thermal_broad}
    \end{equation}
    
\noindent adopting an electron temperature $T_e=10,000$\,K, which implies thermal widths of 10.7, 5.7, and 5.4\,km\,s$^{-1}$ for ions of He, N, and O, respectively ($m_{ion}$ is in atomic mass units).  If the electron temperature is higher (lower) than $T_e=10,000$\,K, the resulting residual velocity will be slightly over(under)-estimated, e.g., if the true electron temperature were 5,000\,K or 15,000\,K instead of the 10,000\,K we adopt, the above thermal widths change by factors of only 0.7 and 1.2.  The effect is very small for the lines of N and O, whose thermal widths are already a small fraction of the typical residual velocity (Figure \ref{fig:vexp_fwhm_errors} or Figure \ref{fig:histograms_fwhm_all_pne}).

Table \ref{tab:longtab_lambdas} presents the results of the previous measurements.  For each object, it includes the velocity splitting and the residual velocities for the main blue- and redshifted components of the nebular shell in each of the emission lines observed.  

For one object, K 3-73, the residual velocity of the blue component of the [\ion{N}{ii}] $\lambda$6583 line was undefined, because its width (FWHM) is less than the broadening due to thermal motions and the instrumental resolution.  Presumably, in this case, the correction for thermal broadening is over-estimated.  The line width matches the corrections for thermal motions and the instrumental resolution if we adopt a temperature of 6,100\,K.  

\subsection{Uncertainties in the velocity splitting and residual velocity}\label{sec_emp_uncert}

\begin{figure}
    \centering
    \includegraphics[width=0.8\columnwidth]{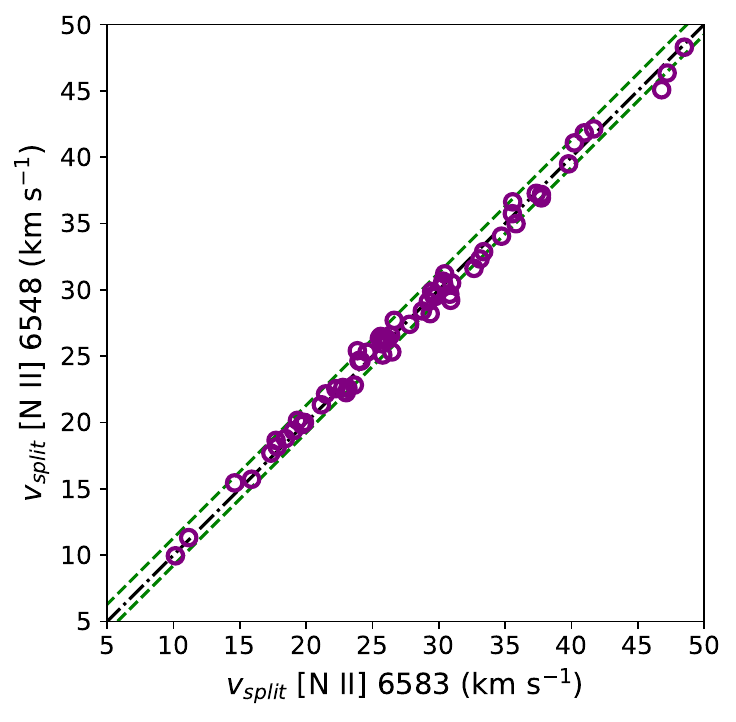}
    \includegraphics[width=0.8\columnwidth]{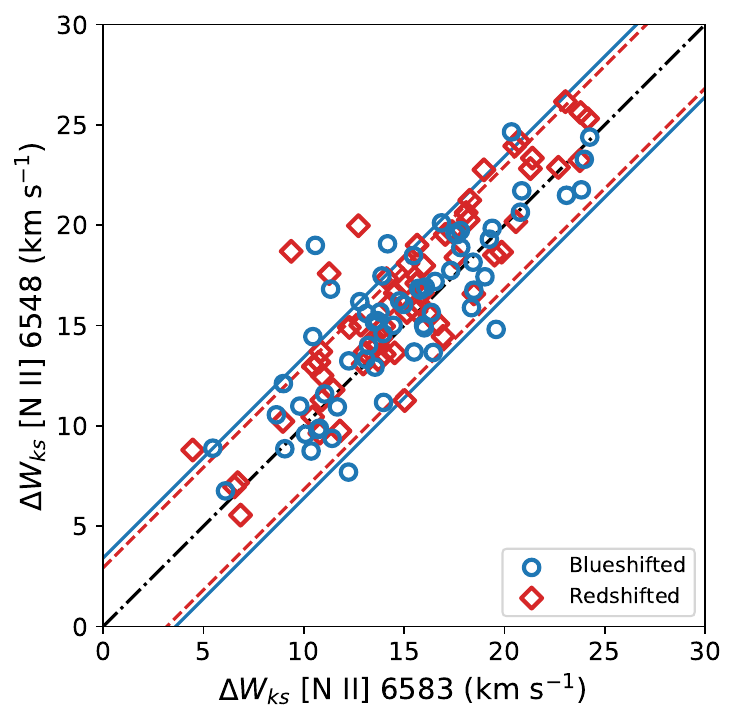}
    \caption{Top panel:  We plot the velocity splitting for the [\ion{N}{ii}] $\lambda$6548 line as a function of that of the [\ion{N}{ii}] $\lambda$6583 line.  The green dashed lines indicate the 5\% and 95\% percentiles of the distribution of differences in the velocity splitting for the two lines, $\pm 1$\,km\,s$^{-1}$, which we adopt as the uncertainty associated with all velocity splittings.  Bottom panel:  We present the residual velocities in the [\ion{N}{ii}] $\lambda$6548 line as a function of those of the [\ion{N}{ii}] $\lambda$6583 line.  The diagonal lines represent the range where the central 90\% of values are included, for the blue- and redshifted sides of the nebular shells separately, $\pm 3.5$\,km\,s$^{-1}$, which we use to define the uncertainty in all residual velocities (blue continuous and red dashed lines, respectively). In both plots, the black dot-dashed lines represent the line of equality.}
    \label{fig:vexp_fwhm_errors}
\end{figure}

\begin{figure}
    \centering
    \includegraphics[width=0.8\linewidth]{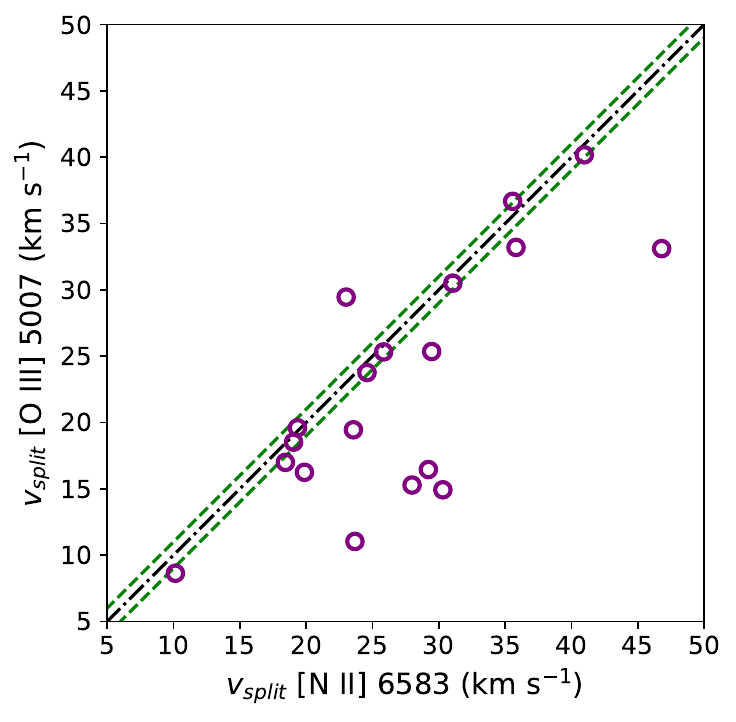}
    \includegraphics[width=0.8\linewidth]{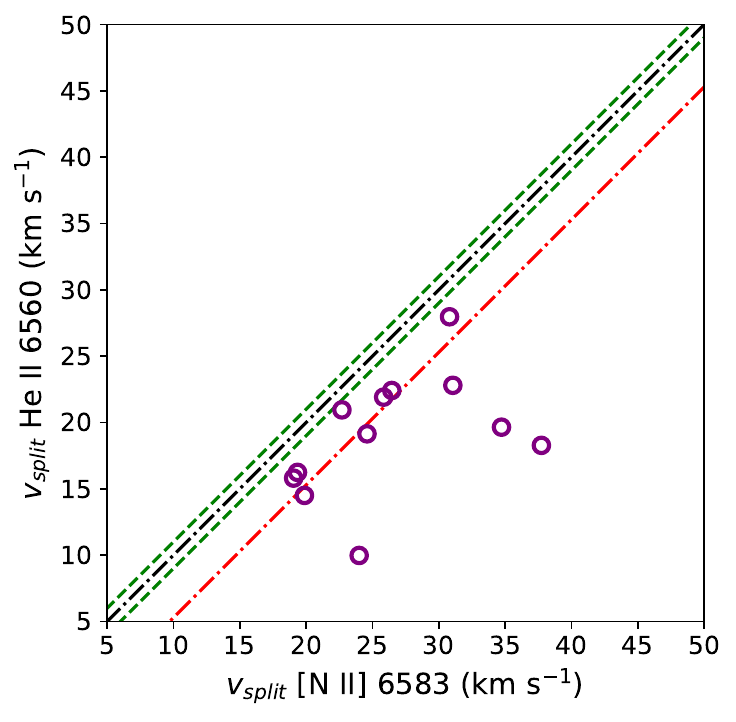}
    \includegraphics[width=0.8\linewidth]{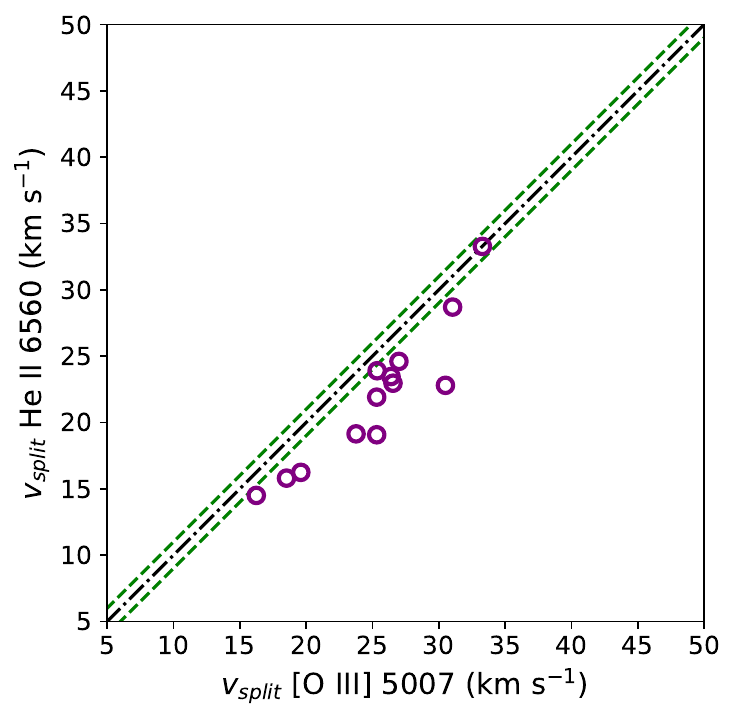}
    \caption{We present pairwise scatter plots of the velocity splitting for the [\ion{N}{ii}], [\ion{O}{iii}] and \ion{He}{ii} lines.  In all panels, the more highly ionized line is on the vertical axis. The green diagonal dashed lines represent the uncertainty about equality, represented by the black dot-dashed line (Figure \ref{fig:vexp_fwhm_errors}).  Most of the points fall near or below these lines in all panels, indicating that the velocity splitting increases for less highly ionized lines.  In the bottom panel, the red line represents the median of the velocity gradients, 4.71\,km\,s$^{-1}$ (\S\ref{sec_charac_Vsplit}).}
    \label{fig:vexp_errors}
\end{figure}

The [\ion{N}{ii}] $\lambda\lambda$6548,6583 lines arise from the same upper level and so should be emitted from exactly the same volume of the nebula and under the same physical conditions.  In the absence of measurement uncertainty, the results for these two lines should agree perfectly.  Hence, we use the degree of coincidence between our measurements for these two lines to estimate the uncertainties in our measurements of the velocity splitting and the residual velocities \citep{richer2017}.  

The top panel in Figure \ref{fig:vexp_fwhm_errors} presents the correlation between the velocity splitting for the [\ion{N}{ii}] $\lambda\lambda$6548,6583 lines.  The distribution of the difference in the velocity splittings for the two lines is centred near zero and the distribution has 5 and 95 percentile points at $-0.91\,\mathrm{km}\,\mathrm{s}^{-1}$ and $+1.15\,\mathrm{km}\,\mathrm{s}^{-1}$, respectively.  Hence, we adopt a typical uncertainty of $\pm 1\,\mathrm{km}\,\mathrm{s}^{-1}$ for the velocity splitting.  

The bottom panel of Figure \ref{fig:vexp_fwhm_errors} presents the same experiment using the residual velocities for the [\ion{N}{ii}] $\lambda\lambda$6548,6583 lines from the blue- and redshifted components.  The distribution of the difference in residual velocities spans a wider range than for the velocity splitting.  For the blue-shifted emission component, the 5 and 95 percentile points are $-4.27\,\mathrm{km}\,\mathrm{s}^{-1}$ and $+2.76\,\mathrm{km}\,\mathrm{s}^{-1}$, respectively.  For the redshifted emission component, the equivalent values are $-4.25\,\mathrm{km}\,\mathrm{s}^{-1}$ to $+1.85\,\mathrm{km}\,\mathrm{s}^{-1}$.  Therefore, for the residual velocities, we adopt an uncertainty of $\pm3.5\,\mathrm{km}\,\mathrm{s}^{-1}$.  

There is also a practical issue to consider, which is the precise positioning of the spectrograph slit over the objects we study.  For the 10 objects for which we have at least two intersecting slits at different position angles, the difference in the residual velocity along the same nominal line of sight (the intersection point) has a flat distribution up to a velocity difference of 5\,km s$^{-1}$.  Presumably, these differences in residual velocities are due to small differences in the precise slit position and the exact micro-structures within the nebular volumes included in each observation.  Hence, our residual velocities should be considered as indicative for that particular nebular shell, rather than an exact value, since it clearly varies within any given object in our sample.

\section{Results}\label{sec_results}

\begin{figure}
    \centering
    \includegraphics[width=0.8\linewidth]{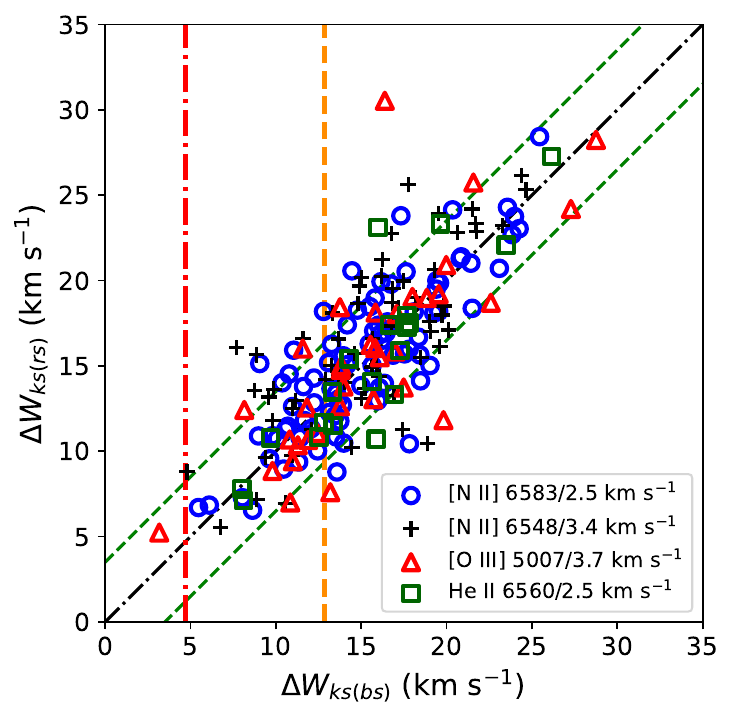}
    \caption{We plot the residual velocities for the receding side of the nebular shell as a function of residual velocity of the approaching side in the four emission lines.  There is a good correlation between the residual velocity measured in a given object and line. The green diagonal dashed lines indicate our estimate of the total uncertainty of the residual velocities (\S\ref{sec_emp_uncert}, Figure \ref{fig:vexp_fwhm_errors}) and the black dot-dashed line indicates equal values. The number after the line wavelength in the legend indicates the rms difference in the residual velocities between the two sides of the nebular shell.  The vertical lines indicate the median velocity gradient through the nebular shells (red dot-dashed; \S\ref{sec_charac_Vsplit}) and the sound speed in a fully-ionized plasma of pure hydrogen at 10,000\,K (orange dashed; see text).}
    \label{fig:histograms_fwhm_all_pne}
\end{figure}

\begin{figure}
    \centering
    \includegraphics[width=0.8\linewidth]{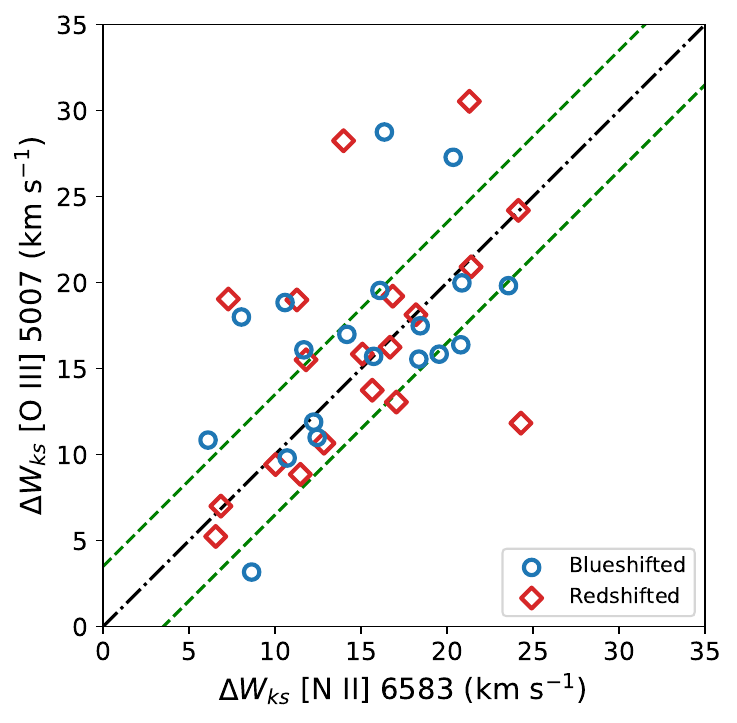}
    \includegraphics[width=0.8\linewidth]{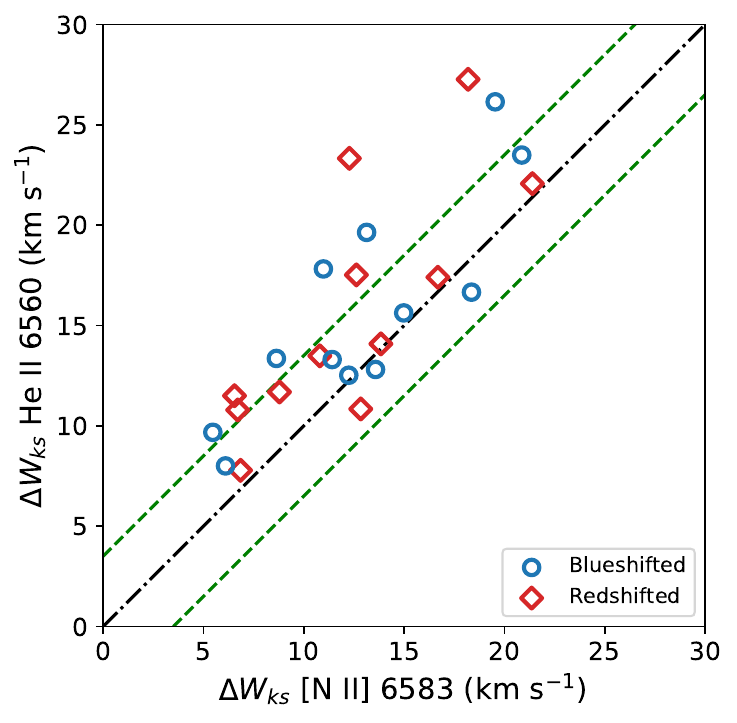}
    \includegraphics[width=0.8\linewidth]{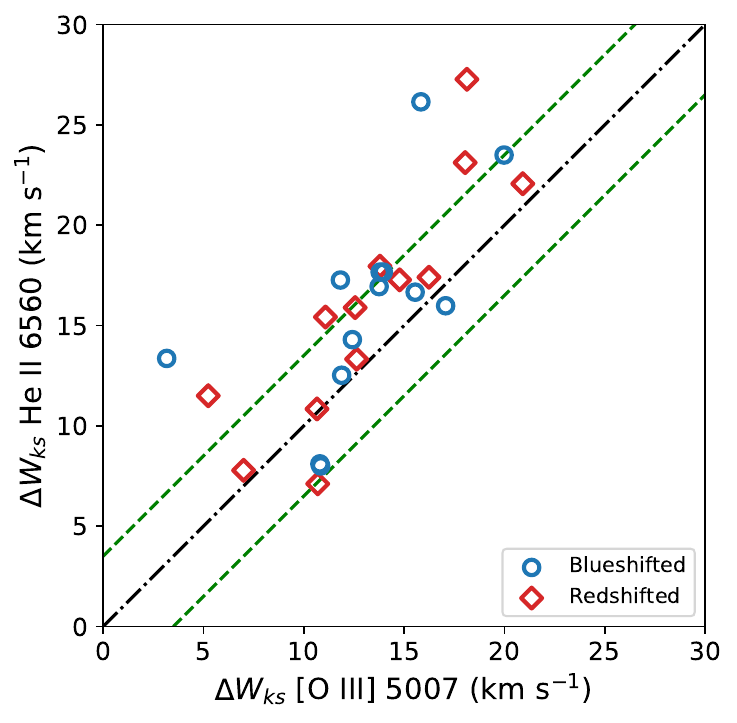}
    \caption{These pairwise scatter plots compare the residual velocities in the [\ion{N}{ii}], [\ion{O}{iii}] and \ion{He}{ii} lines.  The green diagonal dashed lines indicate the uncertainty range about unity for residual velocities and the black dot-dashed indicates equal values.  In all panels, the more highly ionized line is plotted on the vertical axis.  The points clearly scatter above the uncertainty range, indicating that the residual velocity tends to increase for more highly ionized lines, or the more internal parts of the nebular shells.  
    }
    \label{fig:residual_vel_errors}
\end{figure}

\begin{figure*}
    \centering
    \includegraphics[width=0.4\linewidth]{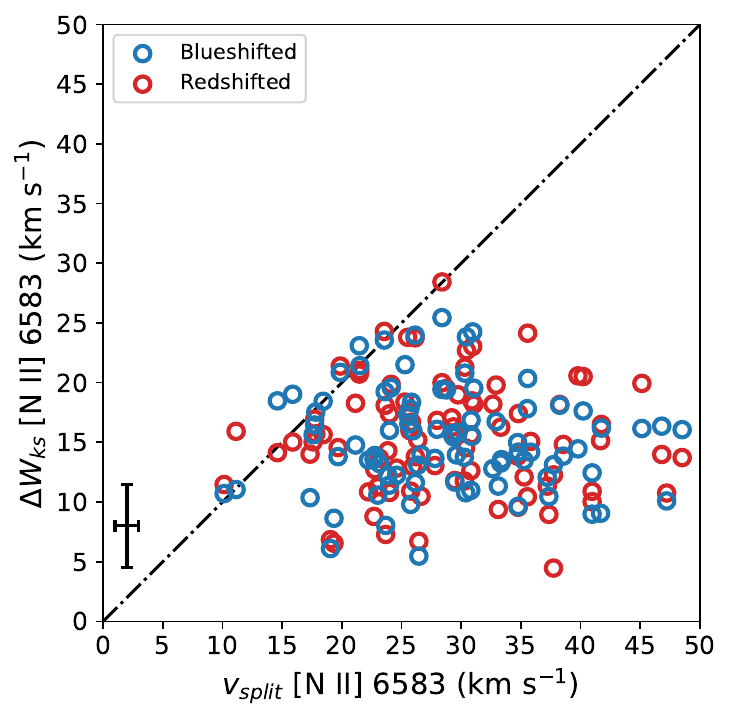}
    \includegraphics[width=0.4\linewidth]{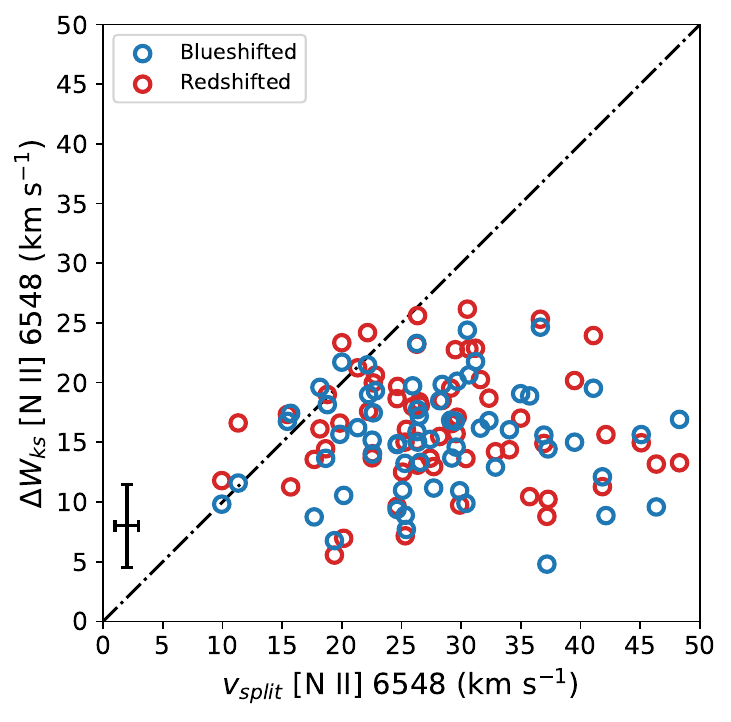}
    \includegraphics[width=0.4\linewidth]{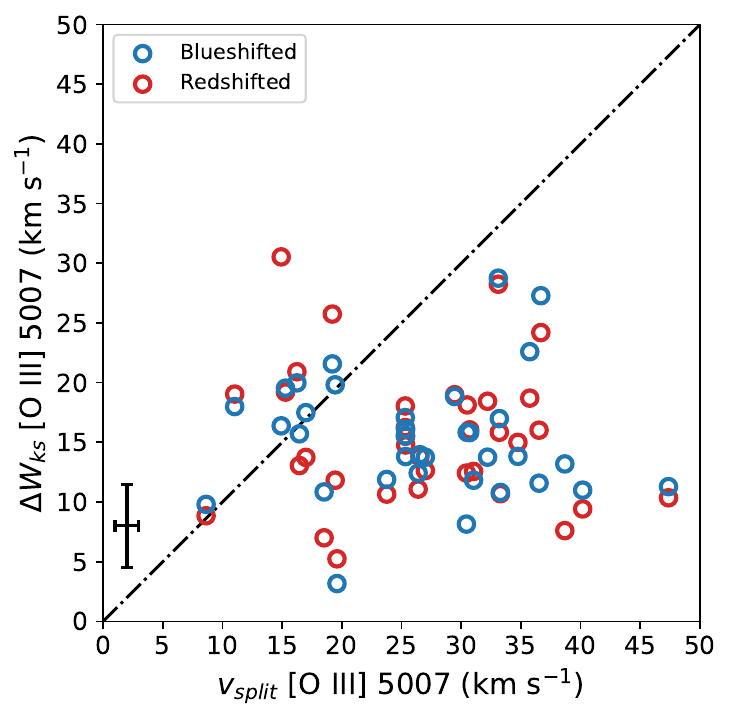}
    \includegraphics[width=0.4\linewidth]{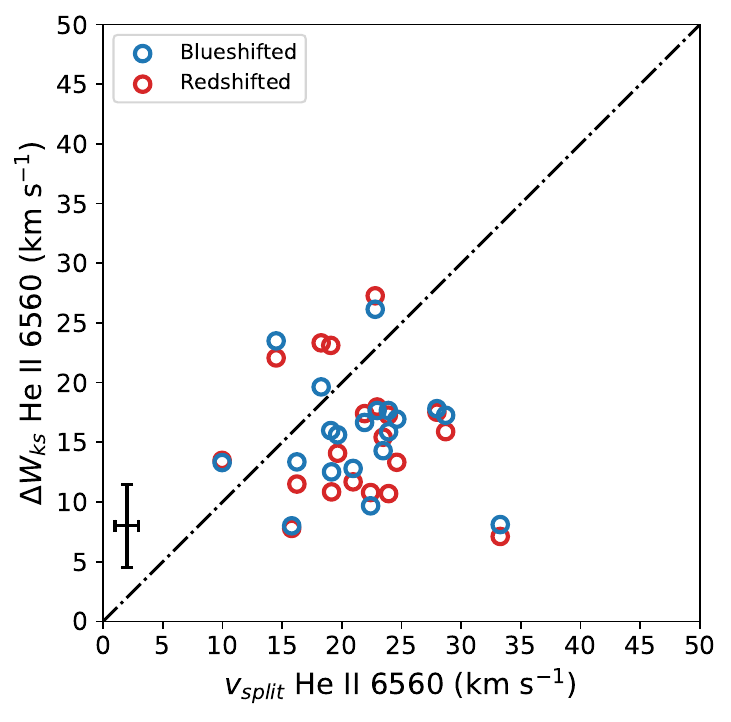}
    \caption{We plot the residual velocity in each  line as a function of the velocity splitting in the same line.  We include the data for the blue- and redshifted sides of the nebular shell, shown in different colours.  The diagonal dot-dashed line indicates the locus of identical values.  Clearly, there is no correlation between the residual velocity and the velocity splitting in any line.  As was found in Figure \ref{fig:histograms_fwhm_all_pne}, the range of residual velocities is similar in all lines.  
    }
    \label{fig:residual_vel_vexp}
\end{figure*}

\subsection{Characteristics of the velocity splitting}\label{sec_charac_Vsplit}

Table \ref{tab:longtab_lambdas} presents the velocity splitting and residual velocities of all the objects in our sample.  As already noted, we use \emph{half} the velocity splitting as defined in Figure \ref{fig:line_prof_fitting}.  The resulting velocity is of a similar magnitude as the expansion velocity of the nebular shell (for the [\ion{N}{ii}] lines).  The top panel in Figure \ref{fig:vexp_fwhm_errors} illustrates that the velocity splitting of the [\ion{N}{ii}] lines spans the range typically observed in planetary nebulae, $10-50$\,km\,s$^{-1}$.  Figure \ref{fig:residual_vel_vexp} presents the same information for the other lines of interest, where we see that the velocity splitting for the [\ion{N}{ii}] and [\ion{O}{iii}] lines span similar ranges, but the \ion{He}{ii} line spans a narrower range.  In Figure \ref{fig:vexp_errors}, it is clear that the usual progression of larger velocity splitting in lower ionization lines occurs generally for the objects in our sample \citep{wilson1950}, since the velocity splitting increases from \ion{He}{ii} $\lambda$6560 to [\ion{O}{iii}] $\lambda$5007 to [\ion{N}{ii}] $\lambda$6563, i.e., usually the data points fall below the band of equal velocities.  

The \ion{He}{ii} and [\ion{N}{ii}] lines sample the inner- and outermost matter in the nebular shell, respectively, so we use the kinematics of these lines to define the velocity difference through the nebular shell.  Given our definition of the velocity splitting, we adopt the difference between the velocity splittings in the [\ion{N}{ii}] and \ion{He}{ii} lines as the velocity difference through the nebular shell.  The middle panel of Figure \ref{fig:vexp_errors} indicates that the velocity gradient can vary substantially, from 1.7 to 19\,km\,s$^{-1}$, but its median value for the 12 objects in our sample where we could observe both lines is 4.7\,km\,s$^{-1}$ (dot-dashed line, middle panel of Figure \ref{fig:vexp_errors}).

\subsection{Characteristics of the residual velocity}

In Figure \ref{fig:histograms_fwhm_all_pne}, we plot the residual velocity for the receding side of the nebular shell for the objects in our sample as a function of the residual velocity of the approaching side for the four emission lines we consider.  It is clear that there is a good correlation between the residual velocity in a given object and emission line.  The diagonal dotted lines denote our estimate of the uncertainty of the residual velocities from \S\ref{sec_emp_uncert} and relatively few objects fall beyond the band defined by these lines.  From Figure \ref{fig:histograms_fwhm_all_pne}, it is also evident that the range of residual velocities in all lines is similar, with the vast majority falling within the range of $7-25$\,km\,s$^{-1}$.  In Figure \ref{fig:histograms_fwhm_all_pne}, we plot the median velocity gradient through the nebular shells from Figure \ref{fig:vexp_errors} as a vertical dot-dashed line.  Almost all of the residual velocities in all lines exceed this velocity gradient.  We also plot the sound speed value for a fully-ionized plasma of pure H at 10,000\,K, $c_s = 12.86$ km s$^{-1}$ as a vertical dashed line.  Considering all emission lines, about 70\% of the residual velocities are supersonic, with only a slight variation of this fraction among the four lines.  

In Figure \ref{fig:residual_vel_errors}, we compare the residual velocities in the [\ion{N}{ii}] $\lambda$6583, [\ion{O}{iii}] $\lambda$5007, and \ion{He}{ii} $\lambda$6560 lines.  In all panels, the more highly ionized line is plotted on the vertical axis.  In contrast to Figure \ref{fig:histograms_fwhm_all_pne}, Figure \ref{fig:residual_vel_errors} compares the residual velocities for these lines on an object-by-object basis.  In this way, it is clear that there is a tendency to find larger residual velocities in more highly ionized lines, i.e., $\Delta W_{ks}$([\ion{N}{ii}])\,$\le \Delta W_{ks}$([\ion{O}{iii}])\,$\le \Delta W_{ks}$(\ion{He}{ii}).  Figure \ref{fig:residual_vel_errors} also emphasizes that the residual velocities in all lines are correlated for a given object.  The residual velocity varies from object-to-object, affecting all lines.  

Figure \ref{fig:residual_vel_vexp} presents the residual velocity as a function of the velocity splitting for the four emission lines.  This figure demonstrates that there is no correlation between the velocity splitting and the residual velocity for any line or, equivalently, in any part of the nebular shell.  There is a similar range of residual velocities at all velocity splittings in all of the lines, as noted in Figure \ref{fig:histograms_fwhm_all_pne}.

\subsection{Correlations with central star properties}

Here, we explore whether the velocity splitting or residual velocity depend upon the properties of the central stars.  The characteristics we consider are the chemical composition, the temperature, and binarity.  We use the nebular excitation as a proxy for the temperature of the central star.  The data required for these analyses are included in Table \ref{tab:longtab_caract}, a review of which quickly reveals that these analyses use only subsets of the full object sample.

The atmospheres of the central stars of planetary nebulae (CSPNe) can be classified as H-rich and H-poor.  However, not all of the data available allow discriminating even this basic characteristic.  We base our analysis upon the compilation from \citet{weidmann2020}.  In our sample, only 43 of the 105 PNe possess a spectral classification in this compilation.  Of these, 25 CSPNe are classified as H-rich (including O(H), DAO, G-K, A, WD, Of-WR(H) and O spectral types), 12 CSPNe are classified as H-poor (9 [WR] and 3 PG1159), with the remaining 6 CSPNe having unclear classifications (wels or continuum).  The wels classification is problematic because the lines used can arise from nebular emission or from irradiation by a close binary companion \citep[see, e.g.,][]{basurah2016}. In what follows, we exclude the last subgroup from consideration. 

\begin{figure}
    \centering
    \includegraphics[width=\linewidth]{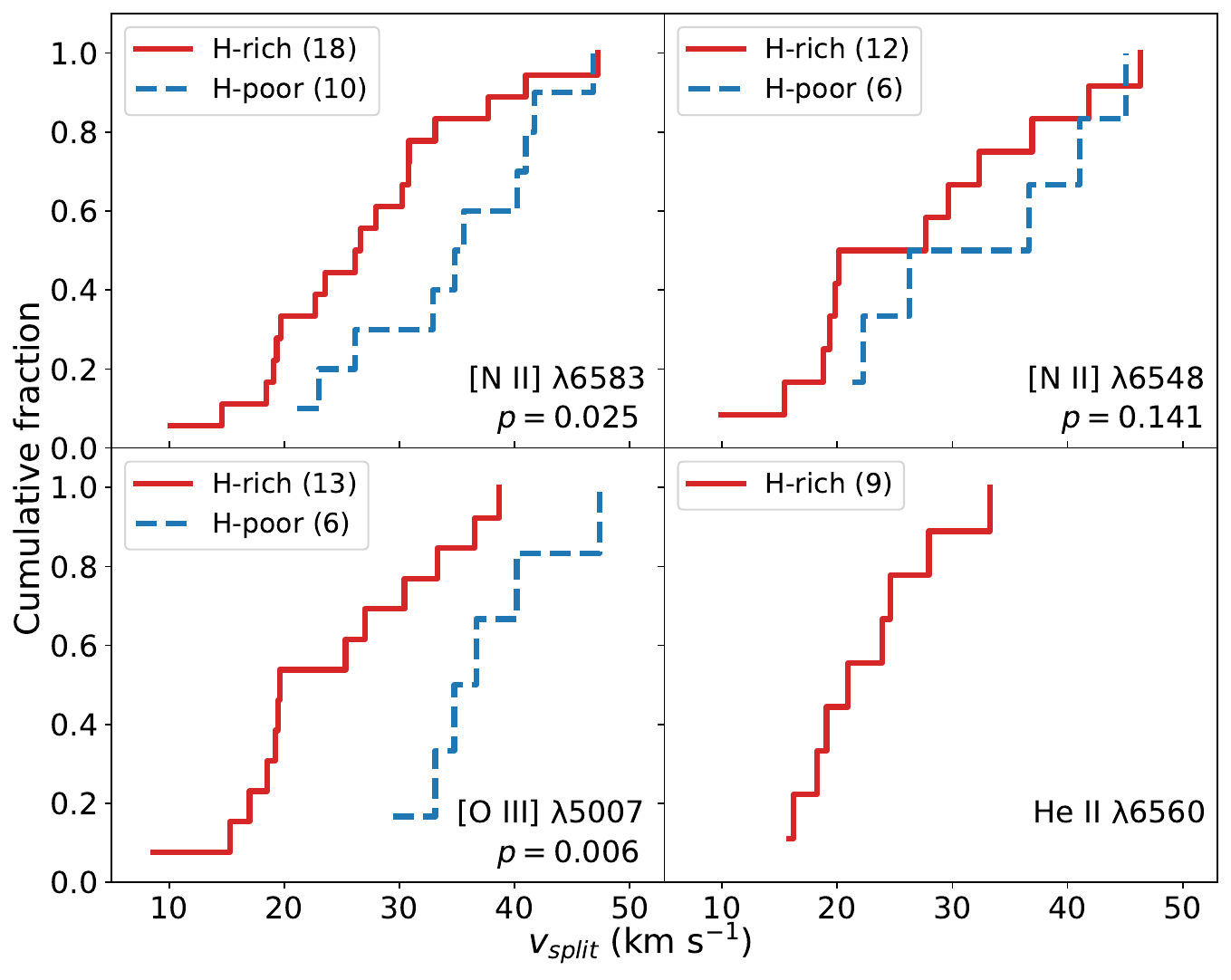}
    \includegraphics[width=\linewidth]{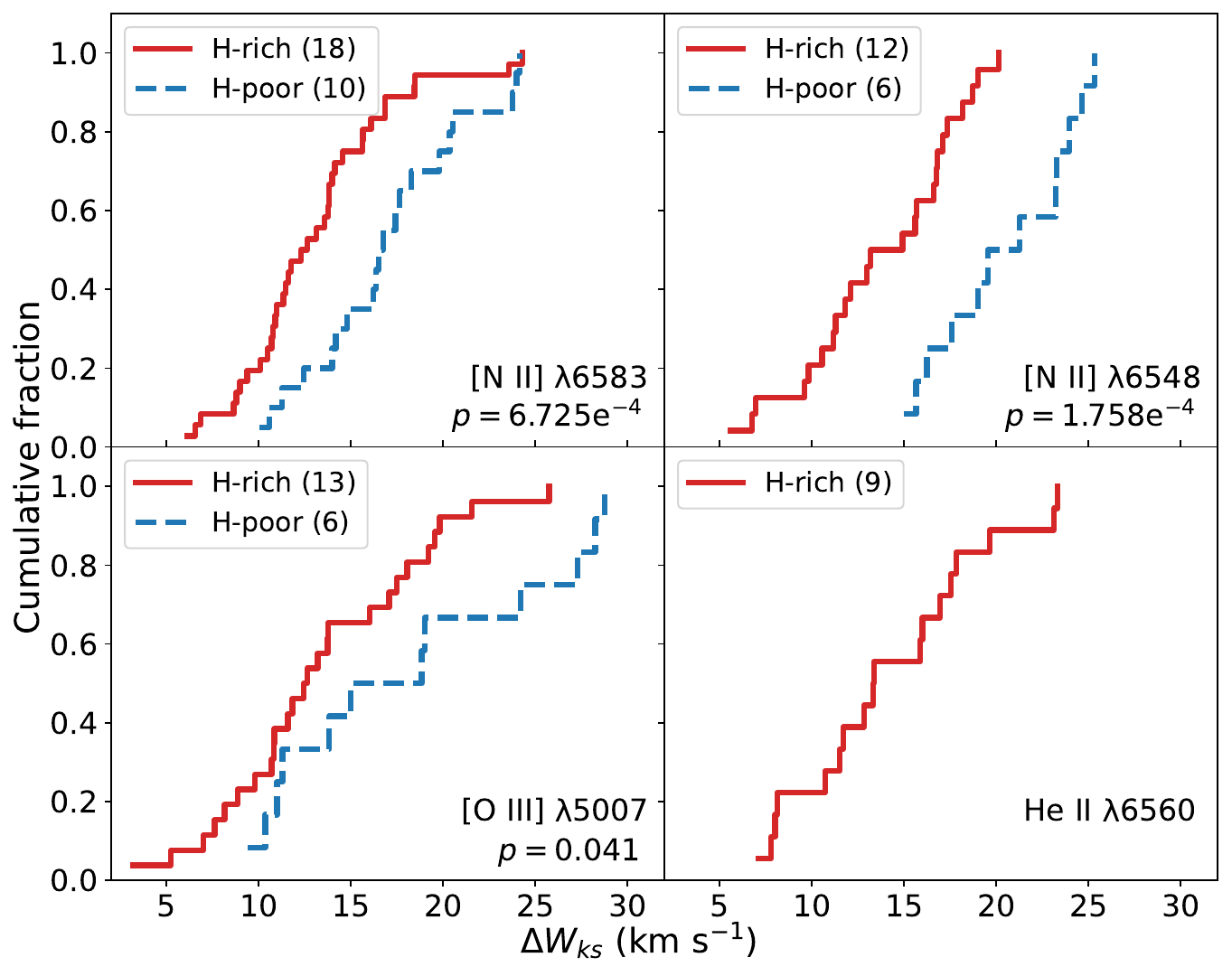}
    \caption{We present the eCDFs for the velocity splitting (top panel) and residual velocities (bottom) of the PNe grouped by H-rich (solid red lines) and H-poor (dashed blue lines) CSPNe.  The number in parentheses indicates the number of objects in each sample.  For the residual velocities we combine those for the approaching and receding sides of the nebular shell.  In each panel, we indicate the p-value for the Wilcoxon-Mann-Whitney U test, considering values below 5\% to indicate a significant difference between the distributions.  By this measure, we find that PNe with H-poor CSPNe have larger velocity splittings and higher residual velocities than their counterparts with H-rich CSPNe.  
    }
    \label{fig:cfds_cspn_atmos}
\end{figure}

In Figure \ref{fig:cfds_cspn_atmos}, we present empirical cumulative distribution functions (eCDFs) of the velocity splitting (top panel) and residual velocities (bottom) for the objects whose central stars may be classified as H-rich and H-poor.  We subdivide each panel to consider each of the four lines of interest individually.  For the residual velocities, we group the results for the approaching and receding sides of the nebula together.  In the H-poor objects, we were able to detect the \ion{He}{ii} line in 7 objects, but it was too faint to obtain reliable results.  The basic result of Figure \ref{fig:cfds_cspn_atmos} is that the objects with H-poor central stars apparently have both larger velocity splitting and larger residual velocities.  

We characterize the significance of these differences using the Wilcoxon-Mann-Whitney U test \citep[henceforth, U-test; e.g.,][]{wall2003} as implemented in the \textsc{Scipy} package \citep{virtanen2020}.  The null hypothesis is that the distributions of for the objects with H-rich and H-poor central stars arise from the same parent distribution.  We test whether the distribution for the objects with H-poor central stars is shifted to larger values (one-tailed test).  The p-value of the test statistic is given in each sub-panel in Figure \ref{fig:cfds_cspn_atmos}.  We adopt the view that p-values below 5\% allow us to reject the null hypothesis and claim that the two distributions are distinct.  

Given the foregoing, Figure \ref{fig:cfds_cspn_atmos} indicates fairly consistent results.  The distributions of the velocity splittings for objects with H-poor central stars are shifted to larger values in both the [\ion{N}{ii}] $\lambda$6583 and [\ion{O}{iii}] $\lambda$5007 lines, but not the [\ion{N}{ii}] $\lambda$6548 line.  While it might seem odd to obtain different results from the [\ion{N}{ii}] lines, the analysis for [\ion{N}{ii}] $\lambda$6548 has substantially fewer objects than that for the [\ion{N}{ii}] $\lambda$6583 line, which may explain the difference.  (We indicate the number of objects in each sample in parentheses in each figure panel.)  As for the residual velocities, the distributions for the objects with the H-poor central stars are shifted to larger values in all of the lines we can compare.  So, rather generally, it seems fair to conclude that planetary nebulae with H-poor central stars exhibit larger velocity splittings and larger residual velocities in all lines than do planetary nebulae with H-rich central stars.  

Figure \ref{fig:cfds_heii} presents the eCDFs for the distributions of velocity splittings and residual velocities according to the degree of excitation of the nebula, which we adopt as a surrogate for the temperature of the central star.  We define objects with central stars that are cooler and hotter, according to whether the \ion{He}{ii} $\lambda$6560 line is absent or present, respectively.  In no line is there a significant difference in the velocity splitting or residual velocity for these different excitation classes.  

Finally, Figure \ref{fig:cfds_cspn_binarity} considers whether the central star is binary.  The primary sources for the information on binarity are \citet{jones2017} and \citet[][see \S\ref{sec_sample_selection} for more details]{bofin2019}.  For the comparison sample, we use the remaining objects from our sample.  While many of these will be single stars, it is very unlikely that none will be binaries.  So, it is very likely that this is a comparison between a sample of binary stars and a mixed sample of single and binary stars.  We find no significant differences between the distributions for the velocity splitting or the residual velocity in these two samples.  Although this result is not conclusive, it is probably fair to conclude that binary central stars do not have a dramatic influence upon either the velocity splitting or the residual velocity.  

For reference, in Table \ref{tab:vexp_fwhm_cspne}, we compile the mean and standard deviations of the velocity splitting and residual velocity in all lines in all the samples used in the foregoing analyses.

\subsection{Correlations with morphology}

\begin{figure}
    \centering
    \includegraphics[width=1\linewidth]{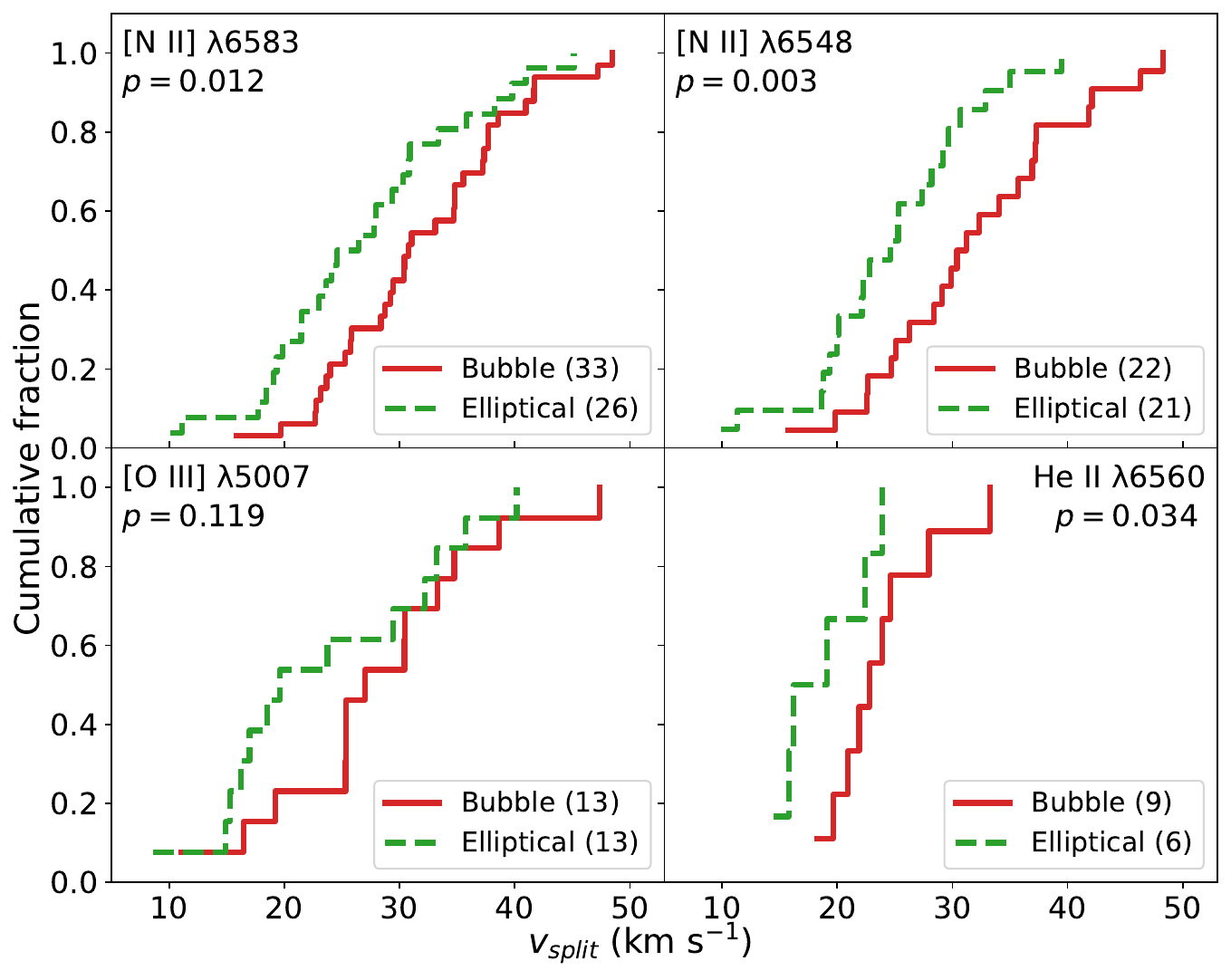}
    \includegraphics[width=1\linewidth]{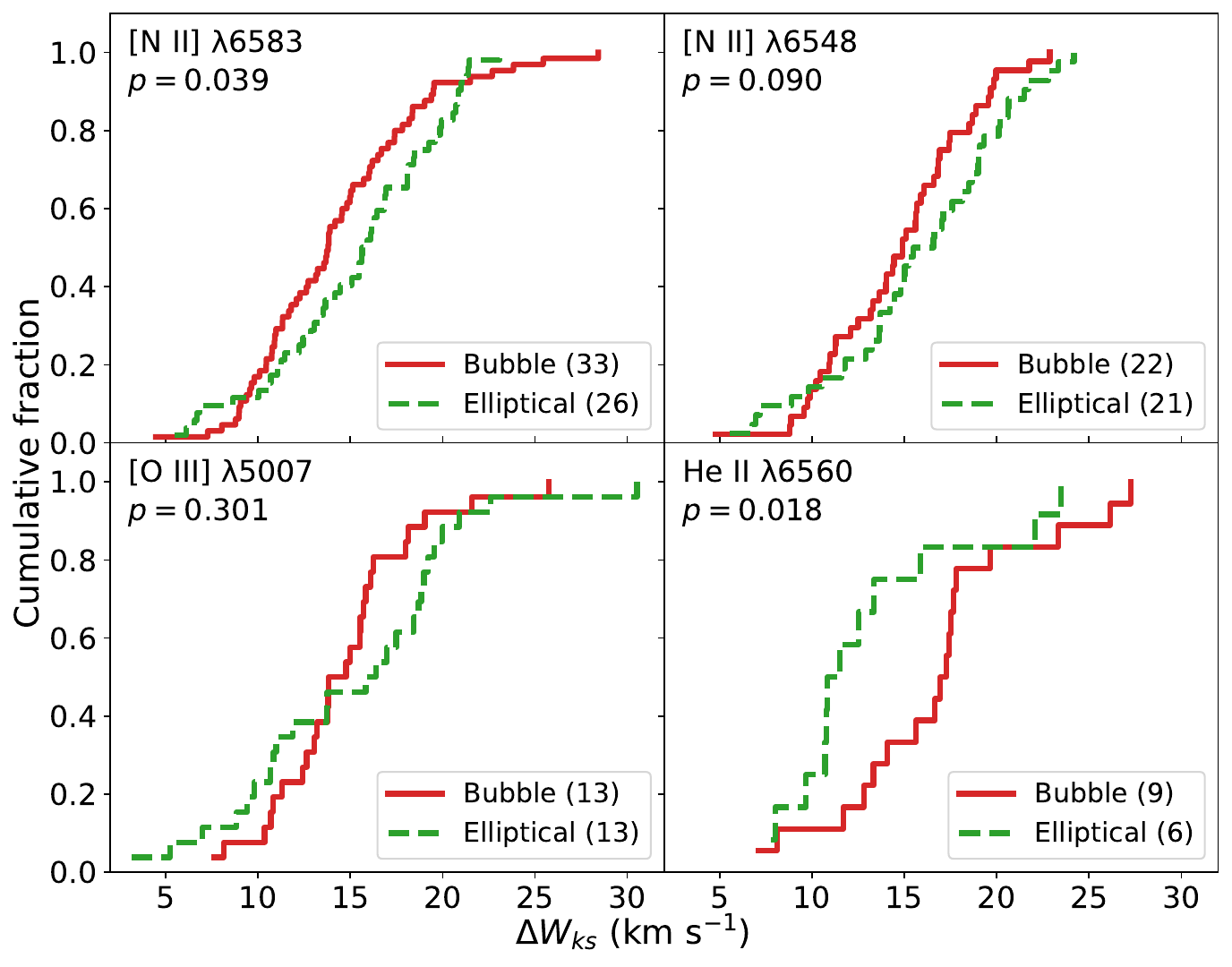}
    \caption{We present the eCDFs compariing the velocity splitting (top panel) and the residual velocities (bottom panel) for PNe with bubble (solid lines) and elliptical (dashed lines) morphologies.  In all lines, the velocity splittings of the PNe with bubble morphology are larger than those of PNe with elliptical morphology, though the difference is not significant for [\ion{O}{iii}] $\lambda$5007.  There are significant differences in the residual velocities in the [\ion{N}{ii}] $\lambda$6583 and \ion{He}{ii} $\lambda$6560 lines, but not consistently for the same morphology.  }
    \label{fig:cfds_morphology}
\end{figure}

Figure \ref{fig:cfds_morphology} compares the eCDFs for objects in our sample with bubble and elliptical morphologies. These two morphologies dominate the sample in all lines (about 63\% overall) and are a direct result of choosing to analyse objects that are spatially-resolved and that have simple morphologies.  Bipolar is the third most common morphology, with only about half the number of elliptical objects.  The toroid and compact morphological types have a similar number of objects each, being each about half as numerous as the bipolar objects.  Finally, there are a few objects with cylindrical morphologies.  The only morphologies with sufficient representation to obtain firm conclusions are those of the bubble and elliptical classes.  Table \ref{tab:vexp_fwhm_morph} provides more details.  

In the top panels of Figure \ref{fig:cfds_morphology}, the distributions of velocity splittings in the [\ion{N}{ii}] $\lambda\lambda$6548,6583 lines and the \ion{He}{ii} $\lambda$6560 line all differ significantly between objects with bubble and elliptical morphologies.  The objects with bubble morphology have systematically larger velocity splitting.  Although this systematic difference also exists for the [\ion{O}{iii}] $\lambda$5007 line, it is not sufficiently large to meet our significance criterion.  So, it seems fair to conclude that generally, objects with bubble morphology expand faster than those with elliptical morphology.  This difference would be expected if the elliptical objects have a prolate shape, which requires that they expand faster towards their poles than towards their equators.  In that case, our choice of the line of sight towards the central star selects an axis along which they expand more slowly.  In contrast, bubble-shaped (round) objects do not suffer such a selection bias.  

In the bottom panels in Figure \ref{fig:cfds_morphology}, the results concerning the residual velocities are mixed.  The [\ion{N}{ii}] $\lambda$6583 and \ion{He}{ii} $\lambda$6560 lines indicate that significant differences exist, but in opposite senses.  In [\ion{N}{ii}] $\lambda$6583, the residual velocities for objects with bubble morphologies are larger, but the elliptical objects have larger residual velocities in \ion{He}{ii} $\lambda$6560.  In the other lines, the differences are not significant.  Thus, as in the case of the binarity of the central star, it is probably fair to say that morphology does not have a strong and clear effect upon the residual velocities.

\subsection{Comparison with past work}\label{sec_comparison_previous}

In comparing our work with previous results, we recall that the typical uncertainty associated with our residual velocities is $\pm 3.5$\,km\,s$^{-1}$ and that similar differences are common even along very similar lines of sight within a given object (\S\ref{sec_emp_uncert}).  Since most previous studies do not provide uncertainties, we assume the same value, except when another value is explicitly given. \citet{guerreroetal1998} studied the shells and haloes of a sample of multiple shell planetary nebulae using data whose spectral resolution matches our highest resolution data, but measured using the H$\alpha$ line, found low residual velocities, $\le 15$\,km\,s$^{-1}$.  Since we do not study outer shells or haloes, we cannot compare with their results.    \citet{acker2002}, based upon their own data and those of \citet{gesickiacker1996}, \citet{gesickietal1996}, and \citet{neineretal2000}, whose spectral resolution is as good or better than ours, and using also the [\ion{N}{ii}] and (one case) [\ion{O}{iii}] lines, find residual velocities systematically less than ours for their planetary nebulae surrounding H-deficient central stars (or wels), though often compatible within our uncertainties.  For none of their planetary nebulae surrounding H-rich central stars do they find a residual velocity, but instead velocity gradients of $20-35$\,km\,s$^{-1}$ that are much larger than our residual velocities.  \citet{gesicki2003} analysed a sample of 73 planetary nebulae using similar data to those above, finding residual velocities only for those with H-deficient central stars.  For the 4 H-deficient objects in common, we typically find somewhat larger residual velocities.  \citet{medina2006} used data of somewhat lower spectral resolution than ours (typical uncertainty of $\pm 8$\,km\,s$^{-1}$) and analysed only the H$\beta$ line without correcting for its fine structure.  They find systematically larger residual velocities than we do for all objects in common, except for NGC 40, for which our value in [\ion{O}{iii}] only, is similar.  Finally, \citet{sabbadin2008} analysed NGC 7009 using a higher spectral resolution (through a 0.9\arcsec slit), finding systematically smaller residual velocities by $4-6$\,km s$^{-1}$ in the \ion{He}{ii} and [\ion{O}{iii}] lines than we do (with our 150$\mu$m slit/1.95\arcsec).  The difference is similar to our uncertainties and also the differences between different observations along the same nominal line of sight (see \S\ref{sec_emp_uncert}).
Therefore, our results are comparable to those from past work when a direct comparison is possible.  

\section{Discussion}\label{sec_discussion}

We begin by summarising our results concerning the residual velocity. We find that residual velocities are ubiquitous, but vary from one object to another. The values we find are typically trans- or supersonic.  Within an object, the residual velocities from different lines are correlated ([\ion{N}{ii}], [\ion{O}{iii}], and \ion{He}{ii}).  The residual velocity does not depend upon the mass of the ion considered, but may increase somewhat as the degree of ionization increases.  There is a systematic difference in residual velocities between planetary nebulae with H-rich and H-poor central stars, with the latter having larger residual velocities. We find no correlation between the residual velocity and the bulk expansion velocity, excitation class, morphology of the nebular shell, or whether the central star is a binary.  

Previous work (\S\ref{sec_comparison_previous}) has clearly established a difference between the nebular shells surrounding H-rich and H-deficient central stars, with the latter clearly showing residual velocities.  Here, we confirm this result, but find that \emph{all} planetary nebulae have a measurable residual velocity.  In part, this is probably due to our selection criterion that requires the line profile to be clearly double-peaked and to the exclusion of \ion{H}{i} lines from our analysis.  An important difference is that our analysis does not depend upon a nebular model, as in \citet{gesickietal1996}, \citet{gesickiacker1996}, and \citet{acker2002}.  Since these models include both velocity gradients and turbulence, it is possible that the two effects are at least partially degenerate in those models.  In fact, that was one of the motivations to analyse only the main emission component in all lines, thereby restricting the volume of the nebular shell involved.  

Our residual velocities span the range of $3-30$\,km\,s$^{-1}$ (Figure \ref{fig:histograms_fwhm_all_pne}), with some 70\% exceeding the sound speed in a pure H plasma at 10,000\,K.  As Figure \ref{fig:histograms_fwhm_all_pne} makes clear, our residual velocities are also larger than the median velocity gradient measured via the [\ion{N}{ii}] and \ion{He}{ii} lines in the nebular shells where they may be measured, though, as stated earlier (\S\ref{sec_sample_selection}), our measurements of the residual velocity should have limited sensitivity to such gradients.  

Both the interacting stellar winds theory \citep{kwoketal1978} and one-dimensional hydrodynamical models \citep[e.g.,][]{perinottoetal2004} predict that the dynamical evolution of the nebular shell is driven by the ionization due to the central star and the energy input from the wind interaction.  Two-dimensional hydrodynamical models further indicate that different instabilities develop very rapidly and have a profound effect upon the evolution of the nebular shell \citep[e.g.,][]{toalaarthur2014,garciaseguraetal2022}.  Figure \ref{fig:residual_vel_errors} illustrates that the residual velocities in a given object are correlated and that they may increase as the ionization increases, since $\Delta W_{ks}(6560)\ge \Delta W_{ks}(5007) \ge \Delta W_{ks}(6583)$.  Both of these results suggest that the wind interaction may be involved in generating the residual velocity since the instabilities will generate disordered macroscopic velocities.  The correlation of the residual velocities in a given object may reflect that the intensity of the instabilities may vary on a case by case basis.  The systematically larger residual velocities observed in the nebulae hosting H-poor central stars (Table \ref{tab:vexp_fwhm_cspne}, Figure \ref{fig:cfds_cspn_atmos}) may also be interpreted as supporting the wind interaction as the origin of the residual velocities \citep{acker2002,gesicki2003,medina2006}.

The result that the residual velocity does not vary as a function of the mass of the ion emitting the line indicates that it cannot result from a thermodynamic process.  Consequently, the residual velocity presumably reflects a macroscopic disordered motion within the plasma.  The most obvious cause to which we could attribute the residual velocity is turbulence.  Recent observations in the infrared sensitive to H$_2$ indicate that the small-scale structure of nebular shells should provide ample opportunity to develop turbulent flows, as these images are dominated by small globules that will interact with the fast stellar wind   \citep{horaetal2006,demarcoetal2022,wessonetal2024,matsuuraetal2025}.  Such small-scale globules easily form \citep{hugginsfrank2006}, as is commonly-found in multidimensional hydrodynamical studies \citep[e.g.,][]{toalaarthur2014,garciaseguraetal2022}.

The residual velocity represents a significant kinetic energy within the nebular plasma.  The residual velocity spans the range of $3-30$\,km s$^{-1}$ (Figure \ref{fig:histograms_fwhm_all_pne}), with a typical value of $15-16$\,km s$^{-1}$ (Table \ref{tab:longtab_lambdas}).  Compared to the thermal velocity of H ions at 10,000\,K, whose thermal velocity is 21.4\,km\,s$^{-1}$, the typical value of the residual velocity represents a kinetic energy of order $(15/21)^2 \sim 50\%$ that of the thermal energy of the plasma.  While the residual velocity does not represent a thermalized energy, it must eventually be thermalized via some dissipative process.  While it is debatable how efficiently the kinetic energy due to the residual velocity may be dissipated, unless the efficiency is zero, the kinetic energy in the residual velocity will contribute to the thermal energy of the nebular plasma.  Hence, the thermal energy of the nebular plasma should exceed that due to photoionization by the central star alone.  

Finally, we checked whether the residual velocities correlate with the abundance discrepancy factor (ADF).  The ADF is the ratio of the abundances of an ion determined using permitted and forbidden emission lines. Its origin has been attributed to temperature fluctuations within a chemical homogeneous plasma \citep[see, e.g.,][]{mendez-delgado2023, peimbert1967} and, for PNe with large ADF values, to the presence of a cold hydrogen-deficient component embedded within a normal warm nebular gas \citep[see, e.g.,][]{garcia-rojas2022, liu2000, richer2022}. In our sample, only 17 objects have a known ADF(O$^{+2}$) value\footnote{\url{https://nebulousresearch.org/adfs/}, last updated on January 2024.} \citep{wesson2018}.  Most of them have values $\leq 5$, and we find no clear correlation between the ADF and the residual velocity.  This is probably not surprising, as the sample is small and biassed to small ADFs.  Thus, we cannot draw a firm conclusion.

\section{Conclusions}\label{sec_conclusions}

In this work, we aimed to study and quantify the residual velocity phenomenon in spatially resolved Galactic planetary nebulae. We performed a detailed and statistical analysis of residual velocities in a sample of 105 objects, the largest sample to the date, based on MES long-slit and high-spectral resolution spectra stored in the SPM Catalogue. These data, together with the selection criteria, allowed us to obtain reliable measurements of residual velocities of the main blue- and redshifted nebular shell components. We focused our analysis on [\ion{N}{ii}], [\ion{O}{iii}] and \ion{He}{ii} lines, which are representative of nebular ionization structure, to examine possible variations of residual velocities through the nebular volume. We aimed to explore the existence of correlations between residual velocities and other nebular and central star properties such as the ionization degree, morphology, central star atmosphere and central star binarity, therefore, our sample is composed of PNe that exhibit a range of these characteristics.

We found that residual velocity is a ubiquitous phenomenon, measured in all lines of interest, regardless of the different nebular and central star properties of each object, but varies from object to object. For PNe in our sample, we quantified residual velocities in a range between 3 -- 30 km s$^{-1}$, being 15 -- 16 km s$^{-1}$ the mean value. About 70\% of these values correspond to trans- and supersonic velocities and, in most cases, are larger than the velocity gradient, quantified from [\ion{N}{ii}] and \ion{He}{ii} velocity splitting difference. The residual velocity represents an important amount of kinetic energy within the nebular plasma, of order 50\%, and is independent of the emitting ion, which indicates that it is not the result of a thermodynamic process.  Therefore, the residual velocity can be interpreted as a dissipative macroscopic non-ordered energy within the nebular plasma, presumably turbulence, which eventually needs to be thermalized at later evolutionary nebular stages. Different instabilities produced by the wind interaction within the nebular shell are responsible for these non-ordered macroscopical velocities. The presence of small-scale structure within the nebular shells and their interaction with the fast stellar winds could produce turbulent flows in the plasma.

We did not find any correlation between residual velocities and the velocity splitting, ionization degree, nebular morphology and binarity of the central star, nonetheless, this is not the case for the ionization structure nor the atmosphere of the central star. For individual objects, we found that residual velocities increase to the inner parts of the nebulae, which is indicated by the $\Delta W_{ks}(6560)\ge \Delta W_{ks}(5007) \ge \Delta W_{ks}(6583)$ tendency; besides, we confirmed that PNe hosting an H-deficient [WR] central stars exhibit larger residual velocities than PNe hosting an H-rich central star, in agreement with previous work \citep[e. g.][]{acker2002, gesicki2003, medina2006}, a result which supports that winds interaction and instabilities scenario as the origin of residual velocities.

Much more work, both from observations and theoretical hydrodynamic models, is still needed to better understand and characterise the residual velocity within PNe.

\section*{Acknowledgements}

This work was supported by Universidad Nacional Autónoma de México Postdoctoral Program (POSDOC), FRE gratefully acknowledges support from his UNAM-DGAPA postdoctoral fellowship. We gratefully acknowledge funding from UNAM-DGAPA grants PAPIIT IN111124 and IG101223. Based upon observations carried out at the Observatorio Astronómico Nacional on the Sierra San Pedro Mártir (OAN-SPM), Baja California, México.  We thank the daytime and night support staff at the OAN-SPM for facilitating and helping obtain our observations (Gabriel Garc\'ia, Francisco Guill\'en, Gustavo Melgoza, Salvador Monrroy, Felipe Montalvo, and Hortensia Riesgo).  We also acknowledge support for the maintenance of the MES from CONACyT (now SECIHTI) and UNAM-DGAPA grants.

Through this work we used the following software: \textsc{iraf} \citep{tody1986, tody1993, fitzpatricketal2024} and the \textsc{python} libraries \textsc{astropy} v. 4.3.1 \citep{astropy2013, astropy2018, astropy2022}, \textsc{matplotlib} v. 3.3.4 \citep{hunter2007}, \textsc{numpy} v. 1.19.3 \citep{harris2020}, \textsc{pandas} v. 1.3.5 \citep{pandas2023} and \textsc{scipy} v. 1.5.3 \citep{virtanen2020}.

\section*{Data Availability}


SPM Catalogue contains the PNe images and slit positions analysed through this work. The analysed fits files spectra will be shared by the corresponding author under reasonable request.



\bibliographystyle{mnras}
\bibliography{references.bib} 







\clearpage

\appendix

\setcounter{table}{0}
\renewcommand{\thetable}{A\arabic{table}}

\section{Supplementary material}

\begin{table*}
\centering
\caption{Number of objects, mean and standard deviation values ($\mu \pm \sigma$) of analysed lines $v_{exp}$ and $\Delta W_{ks}$ grouped by CSPNe atmosphere, binarity and ionization degree. \ion{He}{ii} group includes all PNe in which this line was detected, even if it cannot be measured.  Units in km s$^{-1}$. 
}
\begin{tabular}{lcccccc} \hline
 & H-rich & H-poor & Binary & Unknown & No \ion{He}{ii} & \ion{He}{ii}\\ \hline \hline
\textnormal{[\ion{N}{ii}] $\lambda 6583$} & & & & & &\\ 
\#PNe & 18 & 10 & 14 & 74 & 49 & 39 \\
$v_{exp}$ & 26.62$\pm$9.43 & 34.33$\pm$8.58 & 27.25$\pm$9.97 & 29.11$\pm$7.76 & 28.39$\pm$8.60 & 29.34$\pm$7.53\\
$\Delta W_{ks \, (bs)}$ & 13.34$\pm$4.21 & 16.32$\pm$3.83 & 14.75$\pm$3.57 & 15.24$\pm$4.37 & 14.87$\pm$4.00 & 15.52$\pm$4.54 \\
$\Delta W_{ks \, (rs)}$ & 12.59$\pm$4.12 & 17.56$\pm$4.79 & 13.47$\pm$4.30 & 15.68$\pm$4.62 & 14.86$\pm$4.35 & 15.92$\pm$4.92 \\ \hline
\textnormal{[\ion{N}{ii}] $\lambda 6548$} & & & & \\
\#PNe & 12 & 6 & 8 & 56 & 36 & 28\\
$v_{exp}$ & 26.53$\pm$11.18 & 32.10$\pm$10.17 & 25.23$\pm$11.22 & 28.28$\pm$7.65 & 27.52$\pm$8.65 & 28.39$\pm$7.55 \\
$\Delta W_{ks \, (bs)}$ & 13.59$\pm$4.12 & 19.72$\pm$3.64 & 14.47$\pm$3.59 & 15.52$\pm$4.49 & 14.99$\pm$3.95 & 15.97$\pm$4.89 \\
$\Delta W_{ks \, (rs)}$ & 13.78$\pm$4.37 & 21.04$\pm$4.01 & 14.02$\pm$3.47 & 16.72$\pm$4.86 & 15.74$\pm$4.00 & 17.21$\pm$5.59 \\ \hline
\textnormal{[\ion{O}{iii}] $\lambda 5007$}  & & & & \\
\#PNe & 13 & 6 & 9 & 26 & 8 & 24 \\
$v_{exp}$ & 23.76$\pm$8.97 & 36.92$\pm$6.24 & 28.08$\pm$12.99 & 26.76$\pm$7.67 & 25.97$\pm$13.22 & 25.98$\pm$7.38\\
$\Delta W_{ks \, (bs)}$ & 13.60$\pm$5.30 & 18.50$\pm$7.91 & 13.86$\pm$7.65 & 15.91$\pm$3.89 & 16.44$\pm$5.06 & 15.50$\pm$5.47\\
$\Delta W_{ks \, (rs)}$ & 13.00$\pm$5.67 & 17.70$\pm$7.55 & 12.57$\pm$7.05 & 16.18$\pm$5.14 & 15.13$\pm$3.92 & 16.02$\pm$6.42\\ \hline
\ion{He}{ii} $\lambda 6560$  & & & & \\
\#PNe & 9 & --- & 3 & 17 & --- & 20 \\
$v_{exp}$ & 22.23$\pm$5.79 & --- & 18.94$\pm$2.43 & 21.92$\pm$5.56  & --- & 21.47$\pm$5.28 \\
$\Delta W_{ks \, (bs)}$ & 14.28$\pm$4.10 & --- & 13.93$\pm$1.49 & 15.95$\pm$4.83 & --- & 15.64$\pm$4.52\\
$\Delta W_{ks \, (rs)}$ & 14.01$\pm$6.04 & --- & 12.43$\pm$1.44 & 15.96$\pm$5.70 & --- & 15.43$\pm$5.41\\ \hline
\end{tabular}
\label{tab:vexp_fwhm_cspne}
\end{table*}

\begin{figure}
    \centering
    \includegraphics[width=\linewidth]{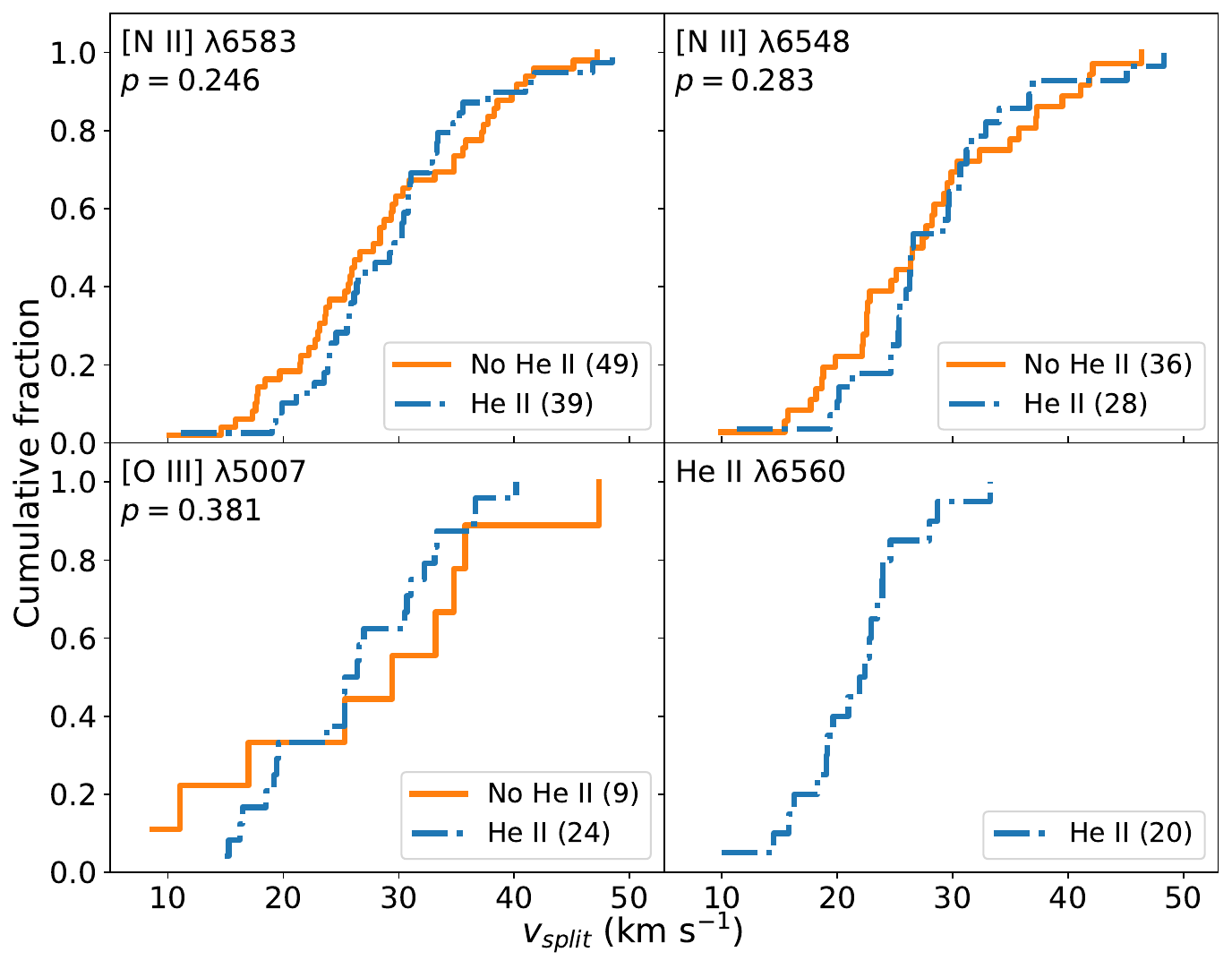}
    \includegraphics[width=\linewidth]{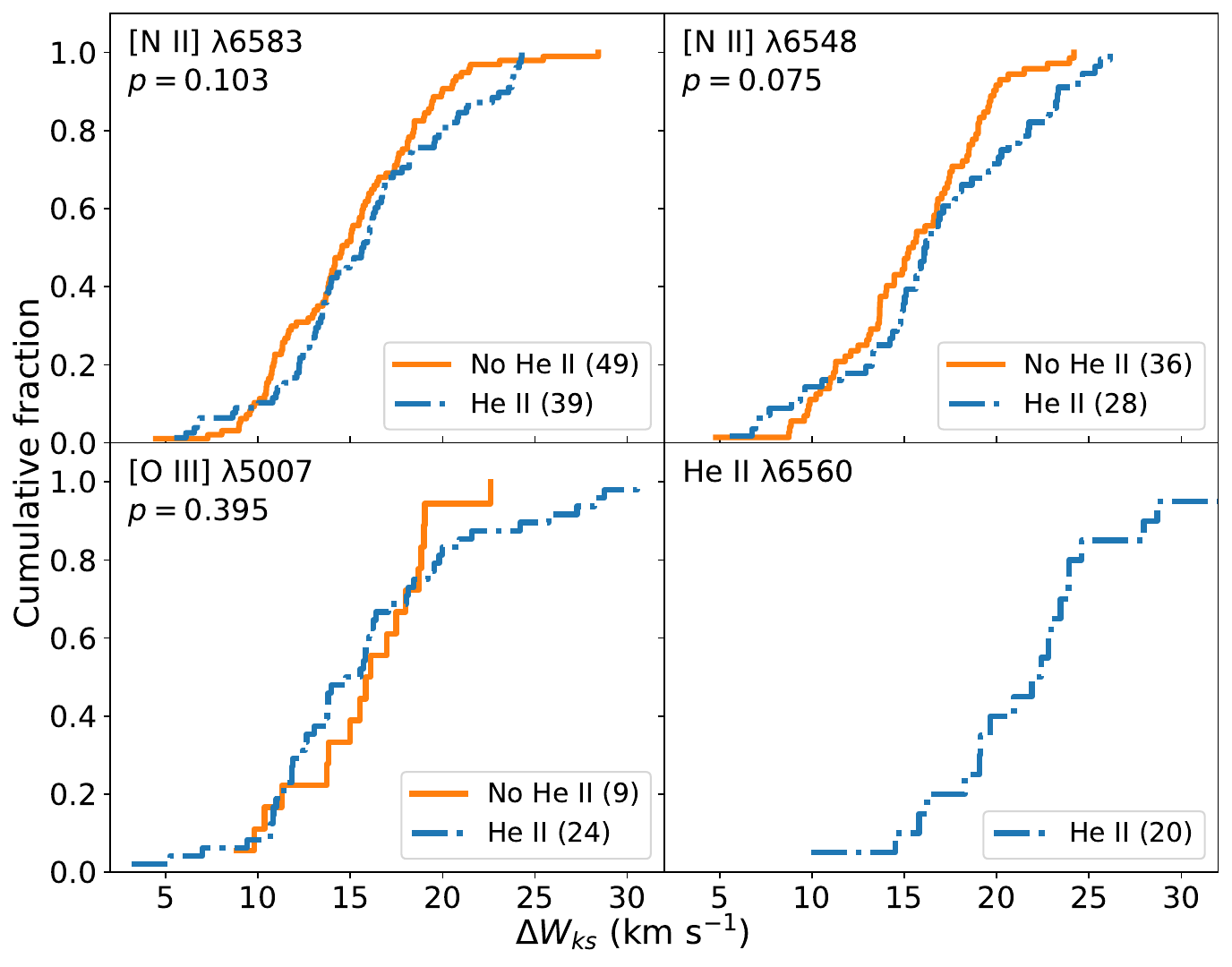}
    \caption{We present eCDFs of the velocity splitting (top panel) and residual velocities (bottom), grouping the objects by the degree of ionization.  We divide the objects into cooler and hotter central stars according to whether the \ion{He}{ii} $\lambda$6560 line is absent or present, respectively.  
    }
    \label{fig:cfds_heii}
\end{figure}

\begin{figure}
    \centering
    \includegraphics[width=\linewidth]{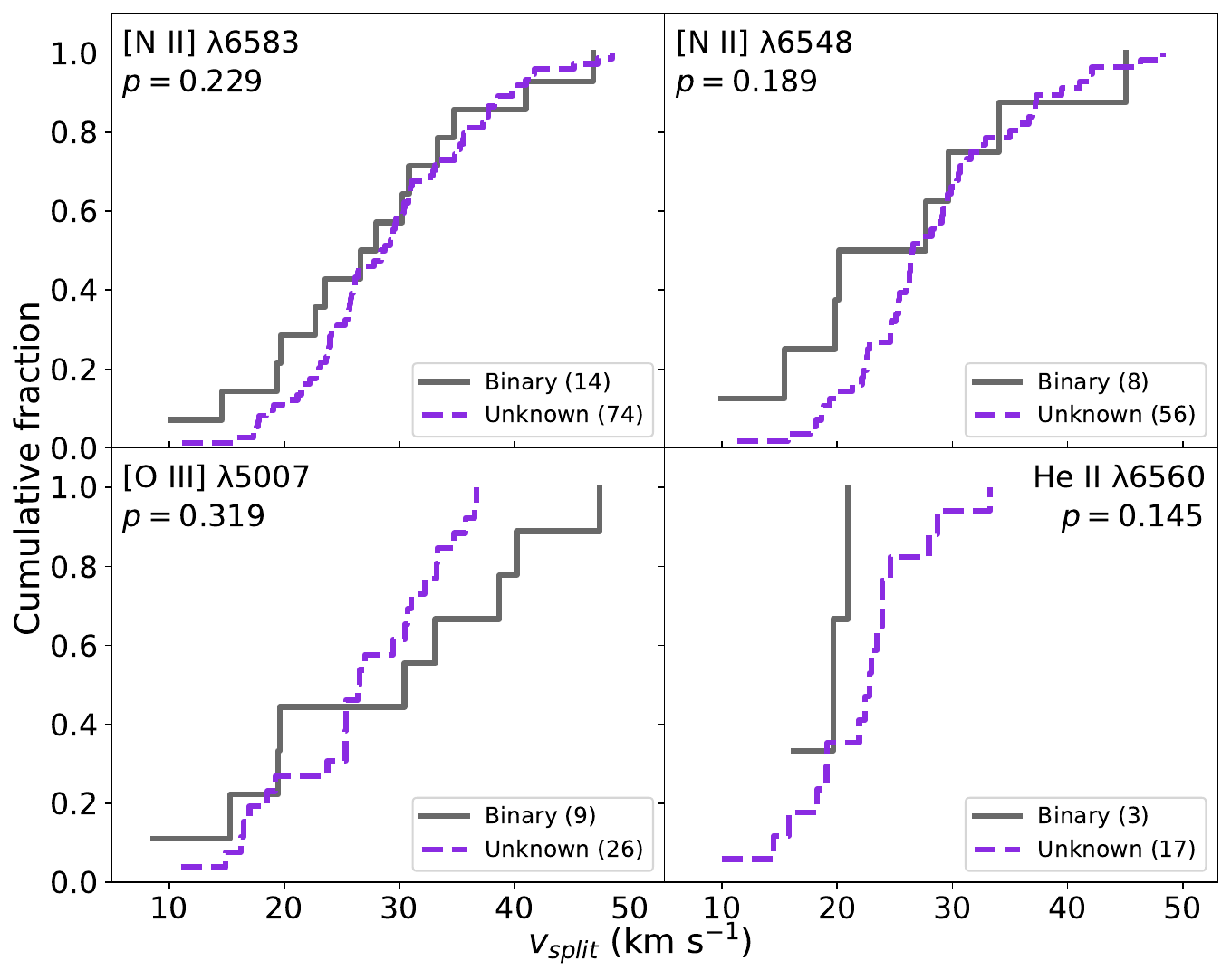}
    \includegraphics[width=\linewidth]{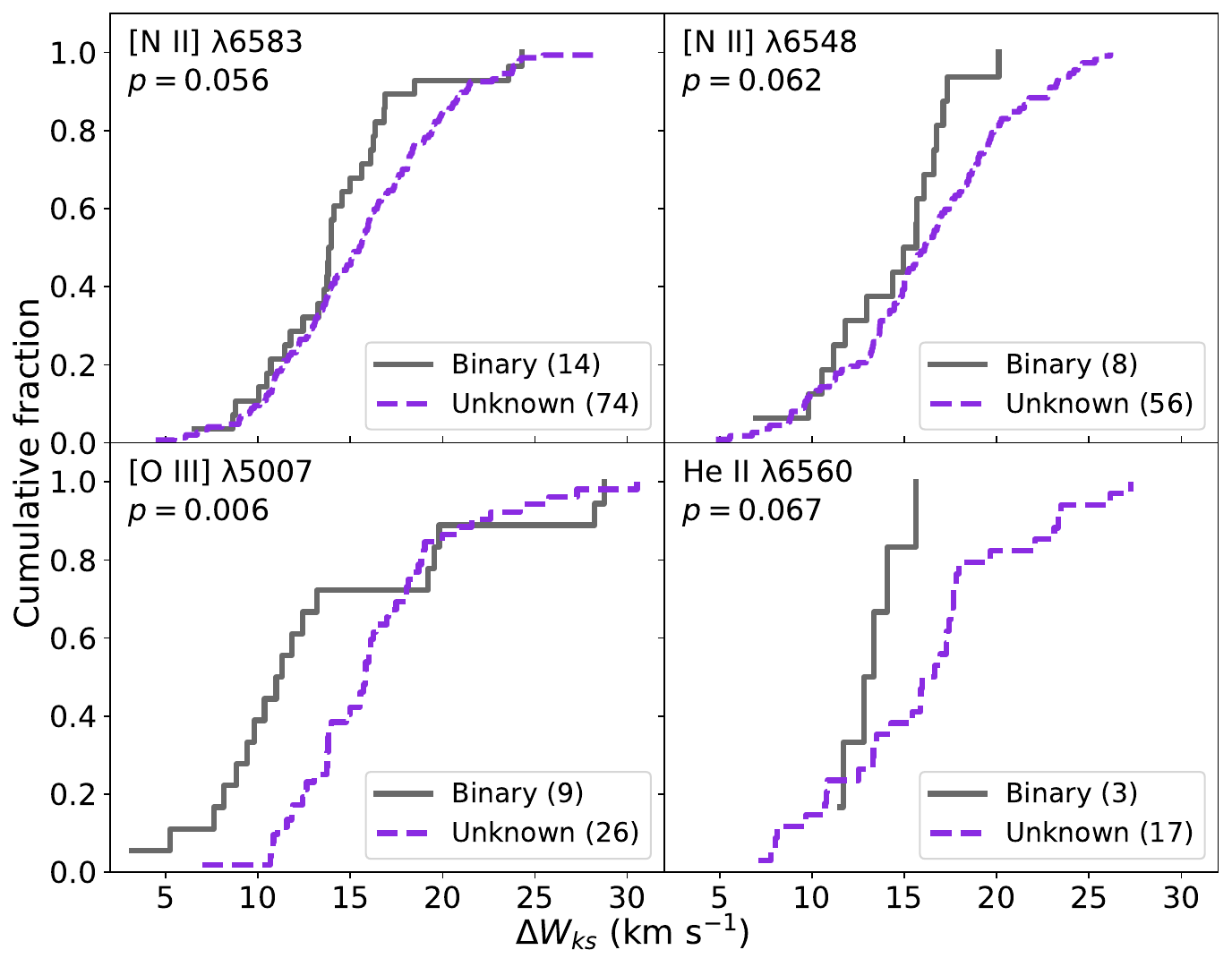}
    \caption{Expansion (left) and residual velocities (right) empirical cumulative distribution functions of analysed lines grouped by binary (solid gray lines) and non-classified as binary (dashed purple lines) CSPNe.
    }
    \label{fig:cfds_cspn_binarity}
\end{figure}

\begin{table*}
\centering
\caption{Number of objects, mean and standard deviation values ($\mu \pm \sigma$) of analysed lines $v_{exp}$ and $\Delta W_{ks}$ grouped by morphology.  Units in km s$^{-1}$.}
\begin{tabular}{lcccccc} \hline
 & Bubble & Elliptical & Bipolar & Toroid & Compact & Cylinder \\ \hline \hline
Total \#PNe & 41 & 29 & 17 & 7 & 7 & 4 \\ \hline
\textnormal{[\ion{N}{ii}]} $\lambda 6583$ &  &  &  &  &  & \\
\#PNe & 33 & 26 & 16 & 7 & 3 & 3 \\
$v_{exp}$ & 31.78$\pm$7.77 & 26.58$\pm$8.82 & 29.00$\pm$7.75 & 27.21$\pm$6.25 & 24.66$\pm$1.28 & 22.47$\pm$6.81 \\
$\Delta W_{ks \, (bs)}$ & 14.45$\pm$4.40 & 15.04$\pm$4.59 & 15.99$\pm$3.26 & 16.86$\pm$3.75 & 15.87$\pm$7.03 & 14.69$\pm$3.48 \\
$\Delta W_{ks \, (rs)}$ & 13.93$\pm$4.65 & 15.54$\pm$4.61 & 17.72$\pm$3.56 & 16.36$\pm$5.13 & 16.28$\pm$6.70 & 12.80$\pm$2.02 \\
\hline
\textnormal{[\ion{N}{ii}]} $\lambda 6548$ &  &  &  &  &  & \\
\#PNe & 22 & 21 & 14 & 2 & 3 & 2 \\
$v_{exp}$ & 31.72$\pm$8.50 & 24.43$\pm$7.23 & 29.62$\pm$7.28 & 20.35$\pm$3.09 & 25.44$\pm$0.83 & 21.57$\pm$8.67 \\
$\Delta W_{ks \, (bs)}$ & 14.53$\pm$4.05 & 15.44$\pm$4.41 & 17.13$\pm$4.13 & 17.39$\pm$3.13 & 13.45$\pm$8.55 & 13.96$\pm$3.95 \\
$\Delta W_{ks \, (rs)}$ & 14.74$\pm$3.84 & 16.1$\pm$5.13 & 19.79$\pm$4.51 & 14.91$\pm$1.72 & 16.31$\pm$6.79 & 15.15$\pm$3.09 \\
\hline
\textnormal{[\ion{O}{iii}]} $\lambda 5007$ &  &  &  &  &  &  \\
\#PNe & 13 & 13 & 3 & 1 & 4 & 1 \\
$v_{exp}$ & 28.05$\pm$9.56 & 23.44$\pm$9.71 & 31.70$\pm$5.81 & 19.44 & 28.67$\pm$2.55 & 36.52 \\
$\Delta W_{ks \, (bs)}$ & 14.43$\pm$3.37 & 14.79$\pm$5.33 & 24.37$\pm$6.36 & 19.82 & 13.51$\pm$1.79 & 11.58 \\
$\Delta W_{ks \, (rs)}$ & 14.71$\pm$4.59 & 15.19$\pm$6.98 & 23.49$\pm$5.14 & 11.83 & 13.37$\pm$2.10 & 16.02 \\ 
\hline
\ion{He}{ii} $\lambda 6560$ &  &  &  &  &  &  \\
\#PNe & 9 & 6 & 1 & --- & 4 & --- \\
$v_{exp}$ & 23.70$\pm$4.59 & 18.67$\pm$3.83 & 19.07 & --- & 21.27$\pm$7.97 & --- \\
$\Delta W_{ks \, (bs)}$ & 16.83$\pm$4.87 & 13.82$\pm$5.49 & 15.99 & --- & 15.63$\pm$2.16 & --- \\
$\Delta W_{ks \, (rs)}$ & 16.56$\pm$6.04 & 12.28$\pm$4.97 & 23.12 & --- & 15.69$\pm$1.83 & --- \\ \hline
\end{tabular}
\label{tab:vexp_fwhm_morph}
\end{table*}

\clearpage 

\section{Long tables}

\onecolumn 

\begin{center}
\begin{footnotesize}

\setcounter{table}{0}
\renewcommand{\thetable}{B\arabic{table}}

\begin{longtable}{llllllll}

\caption{Characteristics of the analysed PNe. Column 1 show the PN G number, Column 2 the common name, Column 3 the filters used during the observations, Column 4 the angular size of the object in \arcsec \citep{acker1992}, Column 5 the morphological classification given in the SPM Catalogue \citep{lopez2012}, Column 6 the CSPNe spectral classification \citep{weidmann2020}, Column 7 the binary identification of CSPNe \citep{jones2017, bofin2019} and Column 8 the emission or absence of \ion{He}{ii} lines. * indicates the objects in which the slit does not crossed over the central star.} 
\label{tab:longtab_caract}
\\

\hline
PN G & Name & Filters & Ang. size (\arcsec) & Morphology & CSPN & Binarity & \ion{He}{ii} emission \\
\hline 
\endfirsthead

\multicolumn{8}{c}%
{{\bfseries \tablename\ \thetable{} -- Continued}} \\
\hline
PN G & Name & Filters & Ang. size (\arcsec) & Morphology & CSPN & Binarity & \ion{He}{ii} emission \\
\hline
\endhead

\hline \multicolumn{8}{r}{{\textit{Continue}}} \\ \hline
\endfoot

\hline
\endlastfoot

000.3$+$06.9 & Terz N 41 & H$\alpha$ & 11 & Elliptical & --- & --- & No \\
000.7$-$03.7 & M 3-22 & H$\alpha$, [\ion{O}{iii}] & 5 & Compact & --- & --- & Yes \\
000.9$-$04.8 & M 3-23 & H$\alpha$, [\ion{O}{iii}] & 11 & Bubble & --- & --- & Yes \\
001.0$+$01.9 & K 1-4 & H$\alpha$ & 37 & Bipolar & --- & --- & No \\
002.6$+$02.1 & K 5-13 & H$\alpha$, [\ion{O}{iii}] & 11 & Elliptical & --- & --- & No \\
003.1$+$04.1 & K 5-7 & H$\alpha$, [\ion{O}{iii}] & 11 & Elliptical & --- & --- & No \\
003.6$-$02.3 & M 2-26 & H$\alpha$, [\ion{O}{iii}] & 9 & Bubble & --- & --- & No \\
004.0$-$05.8 & Pe 1-12 & H$\alpha$, [\ion{O}{iii}] & 10 & Compact & --- & --- & Yes \\
004.8$-$05.0 & M 3-26 & H$\alpha$, [\ion{O}{iii}] & 9 & Bubble & --- & --- & Yes \\
005.1$-$08.9 & Hf 2-2 & H$\alpha$, [\ion{O}{iii}] & 19 & Toroid & O(H)3+? & Binary & Weak \\
005.2$+$05.6 & M 3-12 & H$\alpha$ & 6 & Compact & --- & --- & Weak \\
005.5$-$04.0 & H 2-44 & H$\alpha$, [\ion{O}{iii}] & 7 & Compact & --- & --- & Weak \\
008.8$+$05.2 & Th 4-2 & H$\alpha$ & 19 & Elliptical & --- & --- & Weak \\
010.7$+$07.4 & Sa 2-230 & H$\alpha$, [\ion{O}{iii}] & 10 & Elliptical & --- & --- & Weak \\
011.0$+$06.2 & M 2-15 & H$\alpha$, [\ion{O}{iii}] & 6 & Bubble & --- & --- & No \\
016.0$-$04.3 & M 1-54 & H$\alpha$ & 13 & Bipolar & --- & --- & Weak \\
019.4$-$13.6 & DeHt3 & H$\alpha$ & 32 & Elliptical & --- & --- & No \\
022.5$+$01.0 & MaC 1-13 & H$\alpha$ & 17 & Elliptical & --- & --- & Weak \\
024.2$+$05.9 & M 4-9 & H$\alpha$ & 44 & Bubble & --- & --- & No \\
024.3$-$03.3 & Pe 1-17 & H$\alpha$ & 5 & Bipolar & --- & --- & Weak \\
025.8$-$17.9 & NGC 6818 & H$\alpha$ & 20 & Bubble & wels & Binary & Yes \\
029.2$-$05.9 & NGC 6751 & H$\alpha$ & 21 & Bipolar & [WO4] & --- & No \\
032.7$+$05.6 & K 3-4 & H$\alpha$ & 12 & Bipolar & --- & --- & Weak \\
033.8$-$02.6 & NGC 6741 & H$\alpha$ & 8 & Compact & --- & --- & Yes \\
034.5$-$06.7 & NGC 6778 & H$\alpha$, [\ion{O}{iii}] & 16 & Elliptical & O(H)3-4+? & Binary & Weak \\
036.9$-$01.1 & HaTr11 & H$\alpha$ & 21 & Elliptical & --- & --- & No \\
037.7$-$34.5 & NGC 7009 & H$\alpha$, [\ion{O}{iii}] & 29 & Elliptical & O(H) & --- & Yes \\
040.3$-$00.4 & Abell 53 & H$\alpha$ & 31 & Toroid & --- & --- & No \\
041.8$-$02.9 & NGC 6781 & H$\alpha$ & 108 & Cylinder & DAO & --- & No \\
044.3$-$05.6 & K 3-36 & H$\alpha$, [\ion{O}{iii}] & 12 & Bubble & O? & --- & Weak \\
045.7$-$04.5 & NGC 6804 & H$\alpha$ & 35 & Elliptical & O9 & --- & Yes \\
060.3$-$07.3* & Hen 1-5* & H$\alpha$ & 29 & Bubble & --- & --- & No \\
061.4$-$09.5 & NGC 6905 & H$\alpha$, [\ion{O}{iii}] & 40 & Elliptical & [WO2] & Binary & Weak \\
063.1$+$13.9 & NGC 6720 & H$\alpha$ & 76 & Bubble & hgO(H) & --- & Yes \\
064.6$+$48.2 & NGC 6058 & H$\alpha$, [\ion{O}{iii}] & 23 & Cylinder & O(H)3 & --- & Weak \\
064.9$+$15.5 & M 1-64 & H$\alpha$ & 17 & Bubble & --- & --- & No \\
065.1$-$03.5 & We 1-9 & H$\alpha$ & 24 & Elliptical & --- & --- & No \\
065.2$-$05.6 & Hen 1-6 & H$\alpha$ & 19 & Bipolar & --- & --- & No \\
068.6$+$01.1 & Hen 1-4 & H$\alpha$, [\ion{O}{iii}] & 22 & Elliptical & --- & --- & Weak \\
069.2$+$03.8 & K 3-46 & H$\alpha$ & 23 & Bipolar & --- & --- & No \\
069.4$-$02.6 & NGC 6894 & H$\alpha$ & 40 & Bubble & WD? & --- & No \\
069.7$+$00.0 & K 3-55 & H$\alpha$ & 5 & Bubble & --- & --- & No \\
078.5$+$18.7 & Abell 50 & H$\alpha$ & 27 & Bubble & O(H)III-V & --- & No \\
079.6$+$05.8 & M 4-17 & H$\alpha$ & 15 & Bipolar & --- & --- & Weak \\
083.5$+$12.7 & NGC 6826 & H$\alpha$, [\ion{O}{iii}] & 25 & Elliptical & O(H)3f+? & Binary & No \\
084.0$+$09.5 & K 3-73 & H$\alpha$ & 16 & Bubble & --- & --- & No \\
084.9$-$03.4 & NGC 7027 & H$\alpha$, [\ion{O}{iii}] & 14 & Elliptical & cont & --- & Yes \\
093.3$-$00.9 & K 3-82 & H$\alpha$ & 19 & Bubble & --- & --- & Weak \\
093.3$-$02.4 & M 1-79 & H$\alpha$ & 33 & Bipolar & --- & --- & Weak \\
096.4$+$29.9 & NGC 6543 & H$\alpha$, [\ion{O}{iii}] & 20 & Elliptical & Of-WR(H) & --- & No \\
100.6$-$05.4 & IC 5217 & H$\alpha$ & 7 & Compact & [WC]? & --- & Weak \\
103.2$+$00.6 & M 2-51 & H$\alpha$ & 39 & Bipolar & --- & --- & No \\
104.1$+$07.9 & NGC 7139 & H$\alpha$ & 77 & Bubble & --- & --- & No \\
104.4$-$01.6 & M 2-53 & H$\alpha$ & 15 & Bipolar & WNa? & --- & Weak \\
107.8$+$02.3 & NGC 7354 & H$\alpha$ & 23 & Elliptical & cont & --- & Yes \\
111.2$+$07.0 & KjPn 6 & H$\alpha$ & 6 & Toroid & --- & --- & No \\
112.5$+$03.7 & K 3-88 & H$\alpha$ & 5 & Elliptical & --- & --- & No \\
116.2$+$08.5 & M 2-55 & H$\alpha$ & 39 & Elliptical & --- & --- & Weak \\
118.8$-$74.7 & NGC 246 & H$\alpha$, [\ion{O}{iii}] & 245 & Bubble & PG1159(lgE)+? & Binary & No \\
120.0$+$09.8 & NGC 40 & H$\alpha$, [\ion{O}{iii}] & 48 & Elliptical & [WC8] & --- & No \\
122.1$-$04.9 & Abell 2 & H$\alpha$ & 36 & Bubble & --- & --- & No \\
126.3$+$02.9 & K 3-90 & H$\alpha$, [\ion{O}{iii}] & 9 & Bubble & --- & --- & Yes \\
130.2$+$01.3 & IC 1747 & H$\alpha$ & 13 & Toroid & [WO4] & --- & Weak \\
130.9$-$10.5 & NGC 650-651 & H$\alpha$, [\ion{O}{iii}] & 67 & Bipolar & PG1159(E)+? & Binary & Weak \\
131.4$-$05.4* & BV 5-3* & H$\alpha$ & 24 & Bubble & --- & --- & No \\
138.8$+$02.8 & IC 289 & H$\alpha$, [\ion{O}{iii}] & 35 & Bubble & O(H) & --- & Yes \\
144.1$+$06.1* & NGC 1501* & H$\alpha$ & 52 & Bubble & [WO4] & --- & Weak \\
144.3$-$15.5 & Abell 4 & H$\alpha$ & 20 & Toroid & --- & --- & Weak \\
148.4$+$57.0 & NGC 3587 & H$\alpha$ & 170 & Bubble & DAO & --- & No \\
151.4$+$00.5 & K 3-64 & H$\alpha$ & 8 & Bubble & --- & --- & No \\
171.3$-$25.8* & Ba 1* & H$\alpha$, [\ion{O}{iii}] & 38 & Bubble & [WO]? & --- & No \\
188.6$+$04.4 & HoCr 1 & H$\alpha$ & 61 & Elliptical & --- & --- & No \\
196.6$-$10.9 & NGC 2022 & H$\alpha$ & 19 & Bubble & O(H) & --- & Yes \\
197.2$-$14.2 & K 1-7 & H$\alpha$ & 34 & Bubble & --- & --- & No \\
197.8$-$03.3 & Abell 14 & H$\alpha$ & 33 & Cylinder & A5+? & Binary & No \\
198.6$-$06.3 & Abell 12 & H$\alpha$, [\ion{O}{iii}] & 37 & Bubble & --- & --- & Weak \\
201.9$-$04.6 & We 1-4 & H$\alpha$ & 40 & Bipolar & --- & --- & No \\
206.4$-$40.5 & NGC 1535 & H$\alpha$ & 21 & Bubble & O(H)5+? & Binary & Yes \\
216.0$-$00.2 & Abell 18 & H$\alpha$ & 73 & Bubble & --- & --- & No \\
217.4$+$02.0 & St 3-1 & H$\alpha$ & 15 & Bubble & --- & --- & No \\
221.3$-$12.3 & IC 2165 & H$\alpha$ & 9 & Bubble & wels & --- & Weak \\
231.4$+$04.3* & M 1-18* & H$\alpha$ & 30 & Bubble & --- & --- & No \\
231.8$+$04.1 & NGC 2438 & H$\alpha$ & 64 & Toroid & O(H)+? & Binary & Weak \\
236.7$+$03.5 & K 1-12 & H$\alpha$ & 37 & Elliptical & --- & --- & No \\
238.0$+$34.8 & Abell 33 & [\ion{O}{iii}] & 270 & Bubble & DAO+? & Binary & --- \\
238.9$+$07.3 & Sa 2-21 & H$\alpha$ & 40 & Elliptical & --- & --- & No \\
239.6$+$13.9 & NGC 2610 & H$\alpha$, [\ion{O}{iii}] & 38 & Bubble & WD?+? & --- & Yes \\
243.8$-$37.1 & PRTM 1 & [\ion{O}{iii}] & 19 & Bubble & O(H)3-4 + ? & Binary & --- \\
250.3$+$00.1 & Abell 26 & H$\alpha$ & 40 & Elliptical & --- & --- & No \\
251.1$-$01.5 & K 1-21 & H$\alpha$ & 28 & Elliptical & --- & --- & No \\
252.6$+$04.4 & K 1-1 & H$\alpha$ & 43 & Bubble & G-K+? & Binary & No \\
255.7$+$03.3 & Wray 16-22 & H$\alpha$ & 20 & Bubble & --- & --- & No \\
261.0$+$32.0 & NGC 3242 & H$\alpha$, [\ion{O}{iii}] & 25 & Elliptical & O(H)+? & Binary & Yes \\
261.9$+$08.5 & NGC 2818 & H$\alpha$ & 50 & Bipolar & cont & --- & Weak \\
272.1$+$12.3 & NGC 3132 & H$\alpha$ & 30 & Elliptical & A2V+? & Binary & Weak \\
294.1$+$43.6 & NGC 4361 & H$\alpha$, [\ion{O}{iii}] & 63 & Bipolar & O(H)6 & --- & Yes \\
326.6$+$42.2 & IC 972 & H$\alpha$ & 51 & Bubble & --- & --- & No \\
341.8$+$05.4 & NGC 6153 & H$\alpha$, [\ion{O}{iii}] & 24 & Elliptical & wels & --- & Yes \\
345.9$+$03.0 & Vd 1-6 & H$\alpha$ & 20 & Bubble & --- & --- & No \\
352.9$+$11.4 & K 2-16 & H$\alpha$ & 23 & Bubble & [WC11] & --- & No \\
355.4$-$04.0 & Hf 2-1 & H$\alpha$, [\ion{O}{iii}] & 9 & Bipolar & [WO2] & --- & Weak \\
357.6$-$03.3 & H 2-29 & H$\alpha$ & 5 & Cylinder & O(H)+? & Binary & No \\
358.2$-$01.1 & Bl 1-D & H$\alpha$ & 13 & Toroid & --- & --- & No \\
358.5$+$05.4 & M 3-39 & H$\alpha$ & 18 & Bipolar & --- & Binary & Weak \\
359.4$-$08.5 & Sb 55 & H$\alpha$, [\ion{O}{iii}] & 11 & Compact & --- & --- & Yes \\

\end{longtable}
\end{footnotesize}
\end{center}

\clearpage

\begin{landscape}

\begin{center}
\begin{footnotesize}
\setcounter{table}{1}
\renewcommand{\thetable}{B\arabic{table}}

\setlength{\tabcolsep}{3.5pt} 

\begin{longtable}{lcccccccccccccccccc}

\caption{Velocity splitting and residual velocities (blue- and redshifted) for the analysed PNe grouped by the line of interest, column 1 shows PN G number. All units are in km s$^{-1}$. * indicates the objects in which the slit does not crossed over the central star.}
\label{tab:longtab_lambdas}
\\

\hline
 & \multicolumn{3}{c}{[\ion{N}{ii}] $\lambda 6583$} & \multicolumn{3}{c}{[\ion{N}{ii}] $\lambda 6548$} & \multicolumn{3}{c}{[\ion{O}{iii}] $\lambda 5007$} & \multicolumn{3}{c}{\ion{He}{ii} $\lambda 6560$} & \multicolumn{3}{c}{\ion{C}{ii} $\lambda 6578$} & \multicolumn{2}{c}{\ion{He}{i} $\lambda 5016$} \\
PN G & $v_{split}$ & $\Delta W_{ks (bs)}$ & $\Delta W_{ks (rs)}$ & $v_{split}$ & $\Delta W_{ks (bs)}$ & $\Delta W_{ks (rs)}$ & $v_{split}$ & $\Delta W_{ks (bs)}$ & $\Delta W_{ks (rs)}$ & $v_{split}$ & $\Delta W_{ks (bs)}$ & $\Delta W_{ks (rs)}$ & $v_{split}$ & $\Delta W_{ks (bs)}$ & $\Delta W_{ks (rs)}$ & $v_{split}$ & $\Delta W_{ks (bs)}$ & $\Delta W_{ks (rs)}$
 \\
\hline 
\endfirsthead

\multicolumn{19}{c}%
{{\bfseries \tablename\ \thetable{} -- Continued}} \\
\hline
\hline
 & \multicolumn{3}{c}{[\ion{N}{ii}] $\lambda 6583$} & \multicolumn{3}{c}{[\ion{N}{ii}] $\lambda 6548$} & \multicolumn{3}{c}{[\ion{O}{iii}] $\lambda 5007$} & \multicolumn{3}{c}{\ion{He}{ii} $\lambda 6560$} & \multicolumn{3}{c}{\ion{C}{ii} $\lambda 6578$} & \multicolumn{2}{c}{\ion{He}{i} $\lambda 5016$} \\
PN G & $v_{split}$ & $\Delta W_{ks (bs)}$ & $\Delta W_{ks (rs)}$ & $v_{split}$ & $\Delta W_{ks (bs)}$ & $\Delta W_{ks (rs)}$ & $v_{split}$ & $\Delta W_{ks (bs)}$ & $\Delta W_{ks (rs)}$ & $v_{split}$ & $\Delta W_{ks (bs)}$ & $\Delta W_{ks (rs)}$ & $v_{split}$ & $\Delta W_{ks (bs)}$ & $\Delta W_{ks (rs)}$ & $v_{split}$ & $\Delta W_{ks (bs)}$ & $\Delta W_{ks (rs)}$ \\
\hline
\endhead

\hline \multicolumn{19}{r}{{\textit{Continue}}} \\ \hline
\endfoot

\hline
\endlastfoot

000.3$+$06.9 & 17.73 & 16.45 & 16.94 & 18.65 & 13.66 & 14.46 & --- & --- & --- & --- & --- & --- & --- & --- & --- & --- & --- & --- \\
000.7$-$03.7 & --- & --- & --- & --- & --- & --- & 26.54 & 13.96 & 13.85 & 22.96 & 17.67 & 17.96 & --- & --- & --- & --- & --- & --- \\
000.9$-$04.8 & 31.06 & 19.54 & 18.19 & --- & --- & --- & 30.50 & 15.84 & 18.13 & 22.80 & 26.15 & 27.27 & --- & --- & --- & --- & --- & --- \\
001.0$+$01.9 & 25.61 & 16.54 & 17.58 & 26.48 & 17.20 & 18.40 & --- & --- & --- & --- & --- & --- & --- & --- & --- & --- & --- & --- \\
002.6$+$02.1 & 35.82 & 14.17 & 15.07 & 34.98 & 19.07 & 17.03 & 33.20 & 16.99 & 15.84 & --- & --- & --- & --- & --- & --- & --- & --- & --- \\
003.1$+$04.1 & --- & --- & --- & --- & --- & --- & 35.74 & 22.59 & 18.70 & --- & --- & --- & --- & --- & --- & --- & --- & --- \\
003.6$-$02.3 & 29.46 & 11.67 & 11.80 & 29.87 & 10.94 & 9.75 & 25.35 & 16.09 & 15.51 & --- & --- & --- & --- & --- & --- & --- & --- & --- \\
004.0$-$05.8 & --- & --- & --- & --- & --- & --- & 31.04 & 11.82 & 12.56 & 28.70 & 17.27 & 15.90 & --- & --- & --- & --- & --- & --- \\
004.8$-$05.0 & 25.84 & 18.36 & 16.68 & 26.30 & 15.89 & 15.08 & 25.32 & 15.55 & 16.24 & 21.91 & 16.66 & 17.41 & --- & --- & --- & --- & --- & --- \\
005.1$-$08.9 & 23.56 & 23.57 & 24.29 & --- & --- & --- & 19.44 & 19.82 & 11.82 & --- & --- & --- & --- & --- & --- & --- & --- & --- \\
005.2$+$05.6 & 23.86 & 12.23 & 14.29 & 25.41 & 7.70 & 16.09 & --- & --- & --- & --- & --- & --- & --- & --- & --- & --- & --- & --- \\
005.5$-$04.0 & --- & --- & --- & --- & --- & --- & 30.71 & 15.82 & 16.05 & --- & --- & --- & --- & --- & --- & --- & --- & --- \\
008.8$+$05.2 & 33.38 & 13.55 & 13.40 & 32.87 & 12.93 & 14.21 & --- & --- & --- & --- & --- & --- & --- & --- & --- & --- & --- & --- \\
010.7$+$07.4 & --- & --- & --- & --- & --- & --- & 32.21 & 13.75 & 18.44 & --- & --- & --- & --- & --- & --- & --- & --- & --- \\
011.0$+$06.2 & 23.67 & 8.04 & 7.28 & --- & --- & --- & 11.03 & 18.00 & 19.04 & --- & --- & --- & --- & --- & --- & --- & --- & --- \\
016.0$-$04.3 & 30.96 & 24.25 & 23.04 & 30.51 & 24.39 & 26.16 & --- & --- & --- & --- & --- & --- & --- & --- & --- & --- & --- & --- \\
019.4$-$13.6 & 27.80 & 13.65 & 13.05 & 27.40 & 15.25 & 13.67 & --- & --- & --- & --- & --- & --- & --- & --- & --- & --- & --- & --- \\
022.5$+$01.0 & 11.14 & 11.04 & 15.92 & 11.32 & 11.59 & 16.62 & --- & --- & --- & --- & --- & --- & --- & --- & --- & --- & --- & --- \\
024.2$+$05.9 & 28.39 & 25.45 & 28.43 & --- & --- & --- & --- & --- & --- & --- & --- & --- & --- & --- & --- & --- & --- & --- \\
024.3$-$03.3 & 25.52 & 17.32 & 23.80 & 26.34 & 17.74 & 25.61 & --- & --- & --- & --- & --- & --- & --- & --- & --- & --- & --- & --- \\
025.8$-$17.9 & 34.72 & 14.98 & 13.84 & 34.04 & 16.05 & 14.38 & --- & --- & --- & 19.64 & 15.63 & 14.09 & --- & --- & --- & --- & --- & --- \\
029.2$-$05.9 & 40.21 & 17.63 & 20.51 & 41.07 & 19.53 & 23.95 & --- & --- & --- & --- & --- & --- & --- & --- & --- & --- & --- & --- \\
032.7$+$05.6 & 32.65 & 12.79 & 18.19 & 31.61 & 16.19 & 20.26 & --- & --- & --- & --- & --- & --- & --- & --- & --- & --- & --- & --- \\
033.8$-$02.6 & 23.99 & 11.42 & 10.80 & 24.63 & 9.38 & 9.64 & --- & --- & --- & 9.97 & 13.31 & 13.49 & --- & --- & --- & --- & --- & --- \\
034.5$-$06.7 & 27.98 & 16.09 & 16.84 & --- & --- & --- & 15.28 & 19.54 & 19.21 & --- & --- & --- & --- & --- & --- & 16.81 & 18.61 & 21.28 \\
036.9$-$01.1 & 45.14 & 16.16 & 19.93 & --- & --- & --- & --- & --- & --- & --- & --- & --- & --- & --- & --- & --- & --- & --- \\
037.7$-$34.5 & 19.05 & 6.10 & 6.85 & 19.38 & 6.76 & 5.56 & 18.52 & 10.84 & 7.00 & 15.80 & 8.01 & 7.79 & --- & --- & --- & 18.96 & 13.03 & 8.50 \\
040.3$-$00.4 & 28.39 & 19.43 & 20.00 & --- & --- & --- & --- & --- & --- & --- & --- & --- & --- & --- & --- & --- & --- & --- \\
041.8$-$02.9 & 26.16 & 11.62 & 13.79 & --- & --- & --- & --- & --- & --- & --- & --- & --- & --- & --- & --- & --- & --- & --- \\
044.3$-$05.6 & --- & --- & --- & --- & --- & --- & 19.20 & 21.57 & 25.75 & --- & --- & --- & --- & --- & --- & --- & --- & --- \\
045.7$-$04.5 & --- & --- & --- & --- & --- & --- & --- & --- & --- & 23.92 & 15.88 & 10.71 & --- & --- & --- & --- & --- & --- \\
060.3$-$07.3* & 35.54 & 17.82 & 10.44 & 35.74 & 18.89 & 10.45 & --- & --- & --- & --- & --- & --- & --- & --- & --- & --- & --- & --- \\
061.4$-$09.5 & 40.97 & 12.44 & 10.03 & --- & --- & --- & 40.18 & 10.99 & 9.42 & --- & --- & --- & --- & --- & --- & --- & --- & --- \\
063.1$+$13.9 & 37.73 & 13.13 & 12.27 & 36.93 & 15.60 & 14.91 & --- & --- & --- & 18.27 & 19.64 & 23.33 & --- & --- & --- & --- & --- & --- \\
064.6$+$48.2 & --- & --- & --- & --- & --- & --- & 36.52 & 11.58 & 16.02 & --- & --- & --- & --- & --- & --- & --- & --- & --- \\
064.9$+$15.5 & 25.77 & 9.80 & 10.91 & 25.08 & 10.99 & 12.50 & --- & --- & --- & --- & --- & --- & --- & --- & --- & --- & --- & --- \\
065.1$-$03.5 & 23.63 & 19.25 & 18.09 & 22.82 & 19.31 & 20.63 & --- & --- & --- & --- & --- & --- & --- & --- & --- & --- & --- & --- \\
065.2$-$05.6 & 29.73 & 15.81 & 18.98 & 29.51 & 16.77 & 22.76 & --- & --- & --- & --- & --- & --- & --- & --- & --- & --- & --- & --- \\
068.6$+$01.1 & 30.30 & 20.79 & 21.29 & 30.65 & 20.65 & 22.82 & 14.92 & 16.38 & 30.53 & --- & --- & --- & --- & --- & --- & --- & --- & --- \\
069.2$+$03.8 & 17.34 & 10.36 & 14.01 & 17.69 & 8.75 & 13.57 & --- & --- & --- & --- & --- & --- & --- & --- & --- & --- & --- & --- \\
069.4$-$02.6 & 47.21 & 10.09 & 10.75 & 46.34 & 9.59 & 13.18 & --- & --- & --- & --- & --- & --- & --- & --- & --- & --- & --- & --- \\
069.7$+$00.0 & 23.15 & 13.22 & 13.67 & 22.56 & 14.01 & 14.04 & --- & --- & --- & --- & --- & --- & --- & --- & --- & --- & --- & --- \\
078.5$+$18.7 & 33.11 & 11.34 & 9.38 & 32.32 & 16.81 & 18.70 & --- & --- & --- & --- & --- & --- & --- & --- & --- & --- & --- & --- \\
079.6$+$05.8 & 29.60 & 13.95 & 15.60 & 29.57 & 14.60 & 15.73 & --- & --- & --- & --- & --- & --- & --- & --- & --- & --- & --- & --- \\
083.5$+$12.7 & 10.14 & 10.70 & 11.47 & 9.94 & 9.82 & 11.80 & 8.63 & 9.80 & 8.85 & --- & --- & --- & 7.21 & 10.76 & 10.61 & 8.08 & 8.00 & 8.69 \\
084.0$+$09.5 & 37.73 & --- & 4.47 & 37.18 & 4.80 & 8.80 & --- & --- & --- & --- & --- & --- & --- & --- & --- & --- & --- & --- \\
084.9$-$03.4 & 19.87 & 20.86 & 21.39 & 20.00 & 21.71 & 23.34 & 16.24 & 19.98 & 20.91 & 14.50 & 23.50 & 22.07 & --- & --- & --- & --- & --- & --- \\
093.3$-$00.9 & 48.51 & 16.06 & 13.72 & 48.29 & 16.92 & 13.29 & --- & --- & --- & --- & --- & --- & --- & --- & --- & --- & --- & --- \\
093.3$-$02.4 & 25.70 & 17.81 & 15.98 & 25.93 & 19.73 & 17.98 & --- & --- & --- & --- & --- & --- & --- & --- & --- & --- & --- & --- \\
096.4$+$29.9 & 18.43 & 18.44 & 15.65 & 18.78 & 18.16 & 19.01 & 16.99 & 17.49 & 13.74 & --- & --- & --- & 16.33 & 15.41 & 16.45 & 16.66 & 15.56 & 8.96 \\
100.6$-$05.4 & 26.13 & 23.98 & 23.75 & 26.28 & 23.28 & 23.22 & --- & --- & --- & --- & --- & --- & --- & --- & --- & --- & --- & --- \\
103.2$+$00.6 & 25.95 & 15.97 & 12.96 & 26.34 & 15.02 & 13.08 & --- & --- & --- & --- & --- & --- & --- & --- & --- & --- & --- & --- \\
104.1$+$07.9 & 37.34 & 10.45 & 8.96 & 37.27 & 14.45 & 10.22 & --- & --- & --- & --- & --- & --- & --- & --- & --- & --- & --- & --- \\
104.4$-$01.6 & 21.14 & 14.78 & 18.27 & 21.33 & 16.23 & 21.24 & --- & --- & --- & --- & --- & --- & --- & --- & --- & --- & --- & --- \\
107.8$+$02.3 & 26.45 & 5.47 & 6.71 & 25.31 & 8.89 & 7.17 & --- & --- & --- & 22.41 & 9.68 & 10.79 & --- & --- & --- & --- & --- & --- \\
111.2$+$07.0 & 22.24 & 13.52 & 10.85 & 22.54 & 15.18 & 13.70 & --- & --- & --- & --- & --- & --- & --- & --- & --- & --- & --- & --- \\
112.5$+$03.7 & 21.46 & 23.09 & 20.72 & 22.15 & 21.49 & 24.19 & --- & --- & --- & --- & --- & --- & --- & --- & --- & --- & --- & --- \\
116.2$+$08.5 & 24.13 & 19.59 & 19.86 & 24.63 & 14.81 & 18.66 & --- & --- & --- & --- & --- & --- & --- & --- & --- & --- & --- & --- \\
118.8$-$74.7 & --- & --- & --- & --- & --- & --- & 47.37 & 11.29 & 10.35 & --- & --- & --- & --- & --- & --- & --- & --- & --- \\
120.0$+$09.8 & 23.01 & 10.58 & 11.27 & 22.24 & 19.00 & 17.59 & 29.45 & 18.84 & 18.99 & --- & --- & --- & --- & --- & --- & --- & --- & --- \\
122.1$-$04.9 & 37.21 & 12.08 & 11.33 & --- & --- & --- & --- & --- & --- & --- & --- & --- & --- & --- & --- & --- & --- & --- \\
126.3$+$02.9 & --- & --- & --- & --- & --- & --- & 25.35 & 13.81 & 14.78 & 23.89 & 17.67 & 17.28 & --- & --- & --- & --- & --- & --- \\
130.2$+$01.3 & 32.92 & 16.74 & 19.80 & --- & --- & --- & --- & --- & --- & --- & --- & --- & --- & --- & --- & --- & --- & --- \\
130.9$-$10.5 & 46.80 & 16.35 & 13.98 & 45.08 & 15.65 & 14.96 & 33.11 & 28.75 & 28.24 & --- & --- & --- & --- & --- & --- & --- & --- & --- \\
131.4$-$05.4* & 22.76 & 13.90 & 12.72 & 22.63 & 17.47 & 19.97 & --- & --- & --- & --- & --- & --- & --- & --- & --- & --- & --- & --- \\
138.8$+$02.8 & --- & --- & --- & --- & --- & --- & 26.99 & 13.75 & 12.63 & 24.60 & 16.94 & 13.33 & --- & --- & --- & --- & --- & --- \\
144.1$+$06.1* & 41.74 & 16.19 & 16.50 & --- & --- & --- & --- & --- & --- & --- & --- & --- & --- & --- & --- & --- & --- & --- \\
144.3$-$15.5 & 35.27 & 13.49 & 12.10 & --- & --- & --- & --- & --- & --- & --- & --- & --- & --- & --- & --- & --- & --- & --- \\
148.4$+$57.0 & 40.97 & 8.98 & 10.90 & 41.83 & 12.11 & 11.29 & --- & --- & --- & --- & --- & --- & --- & --- & --- & --- & --- & --- \\
151.4$+$00.5 & 30.37 & 10.78 & 14.53 & 30.40 & 9.88 & 13.64 & --- & --- & --- & --- & --- & --- & --- & --- & --- & --- & --- & --- \\
171.3$-$25.8* & --- & --- & --- & --- & --- & --- & 34.75 & 13.82 & 15.00 & --- & --- & --- & --- & --- & --- & --- & --- & --- \\
188.6$+$04.4 & 30.90 & 15.50 & 18.47 & 29.21 & 13.68 & 16.58 & --- & --- & --- & --- & --- & --- & --- & --- & --- & --- & --- & --- \\
196.6$-$10.9 & 30.81 & 10.98 & 12.62 & --- & --- & --- & --- & --- & --- & 27.97 & 17.82 & 17.52 & --- & --- & --- & --- & --- & --- \\
197.2$-$14.2 & 41.67 & 9.06 & 15.14 & 42.13 & 8.86 & 15.65 & --- & --- & --- & --- & --- & --- & --- & --- & --- & --- & --- & --- \\
197.8$-$03.3 & 14.61 & 18.47 & 14.13 & 15.44 & 16.76 & 17.34 & --- & --- & --- & --- & --- & --- & --- & --- & --- & --- & --- & --- \\
198.6$-$06.3 & 29.21 & 15.72 & 17.05 & 29.12 & 16.85 & 19.54 & 16.45 & 15.71 & 13.05 & --- & --- & --- & --- & --- & --- & --- & --- & --- \\
201.9$-$04.6 & 17.59 & 15.64 & 15.04 & --- & --- & --- & --- & --- & --- & --- & --- & --- & --- & --- & --- & --- & --- & --- \\
206.4$-$40.5 & 22.69 & 13.58 & 8.79 & --- & --- & --- & --- & --- & --- & 20.95 & 12.81 & 11.69 & --- & --- & --- & --- & --- & --- \\
216.0$-$00.2 & 15.88 & 19.03 & 15.03 & 15.72 & 17.43 & 11.26 & --- & --- & --- & --- & --- & --- & --- & --- & --- & --- & --- & --- \\
217.4$+$02.0 & 23.99 & 16.01 & 17.39 & 24.67 & 14.89 & 19.68 & --- & --- & --- & --- & --- & --- & --- & --- & --- & --- & --- & --- \\
221.3$-$12.3 & 30.44 & 23.83 & 22.69 & 31.20 & 21.76 & 22.88 & --- & --- & --- & --- & --- & --- & --- & --- & --- & --- & --- & --- \\
231.4$+$04.3* & 28.75 & 19.39 & 19.50 & 28.41 & 19.85 & 18.51 & --- & --- & --- & --- & --- & --- & --- & --- & --- & --- & --- & --- \\
231.8$+$04.1 & 30.26 & 13.75 & 11.77 & --- & --- & --- & --- & --- & --- & --- & --- & --- & --- & --- & --- & --- & --- & --- \\
236.7$+$03.5 & 21.51 & 21.42 & 21.02 & --- & --- & --- & --- & --- & --- & --- & --- & --- & --- & --- & --- & --- & --- & --- \\
238.0$+$34.8 & --- & --- & --- & --- & --- & --- & 38.68 & 13.19 & 7.61 & --- & --- & --- & --- & --- & --- & --- & --- & --- \\
238.9$+$07.3 & 38.26 & 18.11 & 18.17 & --- & --- & --- & --- & --- & --- & --- & --- & --- & --- & --- & --- & --- & --- & --- \\
239.6$+$13.9 & --- & --- & --- & --- & --- & --- & 33.28 & 10.80 & 10.70 & 33.27 & 8.11 & 7.11 & --- & --- & --- & --- & --- & --- \\
243.8$-$37.1 & --- & --- & --- & --- & --- & --- & 30.44 & 8.15 & 12.42 & --- & --- & --- & --- & --- & --- & --- & --- & --- \\
250.3$+$00.1 & 29.35 & 15.47 & 16.29 & 28.20 & 18.49 & 15.49 & --- & --- & --- & --- & --- & --- & --- & --- & --- & --- & --- & --- \\
251.1$-$01.5 & 39.78 & 14.46 & 20.57 & 39.49 & 15.01 & 20.18 & --- & --- & --- & --- & --- & --- & --- & --- & --- & --- & --- & --- \\
252.6$+$04.4 & 19.69 & 13.80 & 14.56 & 19.84 & 15.67 & 16.61 & --- & --- & --- & --- & --- & --- & --- & --- & --- & --- & --- & --- \\
255.7$+$03.3 & 38.55 & 13.84 & 14.82 & --- & --- & --- & --- & --- & --- & --- & --- & --- & --- & --- & --- & --- & --- & --- \\
261.0$+$32.0 & 19.34 & 8.64 & 6.55 & 20.16 & 10.55 & 6.96 & 19.59 & 3.17 & 5.24 & 16.24 & 13.36 & 11.50 & 19.00 & 12.59 & 7.71 & --- & --- & --- \\
261.9$+$08.5 & 26.36 & 13.09 & 15.19 & 26.57 & 13.32 & 18.10 & --- & --- & --- & --- & --- & --- & --- & --- & --- & --- & --- & --- \\
272.1$+$12.3 & 30.83 & 16.86 & 15.63 & 29.67 & 20.12 & 17.10 & --- & --- & --- & --- & --- & --- & --- & --- & --- & --- & --- & --- \\
294.1$+$43.6 & --- & --- & --- & --- & --- & --- & 25.32 & 17.07 & 18.04 & 19.07 & 15.99 & 23.12 & --- & --- & --- & --- & --- & --- \\
326.6$+$42.2 & 25.29 & 21.50 & 18.38 & --- & --- & --- & --- & --- & --- & --- & --- & --- & --- & --- & --- & --- & --- & --- \\
341.8$+$05.4 & 24.58 & 12.24 & 12.84 & 25.29 & 13.24 & 15.02 & 23.76 & 11.89 & 10.66 & 19.14 & 12.53 & 10.85 & 21.94 & 5.55 & 6.22 & --- & --- & --- \\
345.9$+$03.0 & 34.77 & 9.64 & 9.56 & --- & --- & --- & --- & --- & --- & --- & --- & --- & --- & --- & --- & --- & --- & --- \\
352.9$+$11.4 & 34.79 & 14.18 & 17.41 & --- & --- & --- & --- & --- & --- & --- & --- & --- & --- & --- & --- & --- & --- & --- \\
355.4$-$04.0 & 35.57 & 20.35 & 24.14 & 36.63 & 24.65 & 25.31 & 36.67 & 27.28 & 24.19 & --- & --- & --- & --- & --- & --- & --- & --- & --- \\
357.6$-$03.3 & 26.64 & 13.97 & 10.48 & 27.70 & 11.17 & 12.97 & --- & --- & --- & --- & --- & --- & --- & --- & --- & --- & --- & --- \\
358.2$-$01.1 & 17.82 & 17.54 & 15.71 & 18.17 & 19.61 & 16.13 & --- & --- & --- & --- & --- & --- & --- & --- & --- & --- & --- & --- \\
358.5$+$05.4 & 33.31 & 13.27 & 16.27 & --- & --- & --- & --- & --- & --- & --- & --- & --- & --- & --- & --- & --- & --- & --- \\
359.4$-$08.5 & --- & --- & --- & --- & --- & --- & 26.39 & 12.42 & 11.08 & 23.46 & 14.30 & 15.43 & --- & --- & --- & --- & --- & --- \\

\end{longtable}

\end{footnotesize}
\end{center}
\end{landscape}

\bsp	
\label{lastpage}

\end{document}